\RequirePackage{lineno}
\documentclass[aps,prd,nofootinbib,floats,floatfix,amsfonts,amssymb,amsmath,notitlepage,superscriptaddress]{revtex4-1}
\usepackage{graphicx,color,units,bm,subfigure}
\usepackage{dcolumn}
\usepackage{isomath}
\usepackage[protrusion=true,tracking=true,kerning=true,spacing=true,final,babel=true]{microtype}
\usepackage{physics}
\usepackage{xcolor}
\usepackage{color}
\usepackage{natbib}
\usepackage{graphicx}
\usepackage{ulem}
\usepackage{soul}
\usepackage{multirow}

\begin{document}
\title{Gravitational background from dynamical binaries and detectability with 2G detector}

\date{\today}



\author{Carole~P\'{e}rigois}
\email{\texttt{caroleperigois@outlook.com}}
 \affiliation{LAPP, CNRS, 9 Chemin de Bellevue, 74941 Annecy-le-Vieux, France}
 
\author{Filippo~Santoliquido}%
\affiliation{%
 Physics and Astronomy Department Galileo Galilei, University of Padova, Vicolo dell’Osservatorio 3, I–35122, Padova, Italy}\affiliation{INFN - Padova, Via Marzolo 8, I–35131 Padova, Italy}\affiliation{INAF - Osservatorio Astronomico di Padova, Vicolo dell’Osservatorio 5, I-35122 Padova, Italy}
 
\author{Yann~Bouffanais}%
\affiliation{%
 Physics and Astronomy Department Galileo Galilei, University of Padova, Vicolo dell’Osservatorio 3, I–35122, Padova, Italy}\affiliation{INFN - Padova, Via Marzolo 8, I–35131 Padova, Italy}\affiliation{INAF - Osservatorio Astronomico di Padova, Vicolo dell’Osservatorio 5, I-35122 Padova, Italy}
 
\author{Ugo~N. Di Carlo}%
\affiliation{%
 Physics and Astronomy Department Galileo Galilei, University of Padova, Vicolo dell’Osservatorio 3, I–35122, Padova, Italy}\affiliation{INFN - Padova, Via Marzolo 8, I–35131 Padova, Italy}\affiliation{INAF - Osservatorio Astronomico di Padova, Vicolo dell’Osservatorio 5, I-35122 Padova, Italy}

\author{Nicola~Giacobbo}%
\affiliation{%
 Physics and Astronomy Department Galileo Galilei, University of Padova, Vicolo dell’Osservatorio 3, I–35122, Padova, Italy}\affiliation{School of Physics and Astronomy \& Institute for Gravitational Wave Astronomy, University of Birmingham, Birmingham, B15 2TT, UK}
 
\author{Sara~Rastello}%
\affiliation{%
 Physics and Astronomy Department Galileo Galilei, University of Padova, Vicolo dell’Osservatorio 3, I–35122, Padova, Italy}\affiliation{INFN - Padova, Via Marzolo 8, I–35131 Padova, Italy}
 
 \author{Michela~Mapelli}%
\affiliation{%
 Physics and Astronomy Department Galileo Galilei, University of Padova, Vicolo dell’Osservatorio 3, I–35122, Padova, Italy}\affiliation{INFN - Padova, Via Marzolo 8, I–35131 Padova, Italy}\affiliation{INAF - Osservatorio Astronomico di Padova, Vicolo dell’Osservatorio 5, I-35122 Padova, Italy}
 
 \author{Tania~Regimbau}
 \affiliation{LAPP, CNRS, 9 Chemin de Bellevue, 74941 Annecy-le-Vieux, France}
 
\begin{abstract}
We study the impact of young clusters on the gravitational wave background from compact binary coalescence. We simulate a catalog of sources from population I/II isolated binary stars and stars born in young clusters, corresponding to one year of observations with second-generation (2G) detectors. 
Taking into account uncertainties on the fraction of dynamical binaries and star formation parameters, we find that the background is dominated by the population of binary black holes, and we obtain a value of $\Omega_{gw}(25 \rm{Hz}) =  1.2^{+1.38}_{-0.65} \times 10^{-9}$ for the energy density, in agreement with the actual upper limits derived from the latest observation run of LIGO--Virgo. 
We demonstrate that a large number of sources in a specific corrected mass range yields to a bump in the background.
This background could be detected with 8 years of coincident data by a network of 2G detectors. 
\end{abstract}
\maketitle
\section{Introduction}
Gravitational wave (GW) astronomy started in September 2015 with the first detection of the merger of a binary black hole (BBH)\cite{1602.03838} 
by the two American detectors Advanced LIGO (aLIGO\cite{aLIGO}). In August 2017,  after the  detector Advanced Virgo (AdVirgo \cite{adVirgo}) 
joined the network, the first observation of the coalescence of a binary neutron star (BNS)  \cite{1710.05832, 1710.05833}, in coincidence with electromagnetic counterparts, brought GW astronomy in the multi-messenger era. After a few years and two upgrades of the detectors, aLIGO and AdVirgo have released an updated catalog of 50 events, with 39 new events detected in the first half of the the third observation run (O3a). 

The compact binary coalescences (CBCs) that we are detecting now are loud and close events, suggesting that  a larger number of unresolved sources at higher redshift, too faint to be detected individually, combine to create a background of GWs. This background has  been intensively investigated in the past ~\cite{1101.2762, 1104.3565, 1106.5795, 1111.6125, 1112.1898, 1209.0595, 1205.4621} and predicted to be detected a few years after the second generation (2G) detectors have reached their design sensitivities. In most studies, only CBCs formed through isolated evolution of population I/II binary stars have been considered (hereafter isolated binaries).  In a recent work, including population III stars, Périgois et al.(2021, \cite{2008.04890}) have shown that isolated binaries from population I/II stars are the main contribution in 2G detectors but that population  III stars could dominate the residual background, after detected sources have been removed, in 3G detectors, increasing the amplitude and modifying the shape of the spectrum at low frequencies.

One of the new events detected during O3a, the BBH merger GW190521 \cite{2020PhRvL.125j1102A,2020ApJ...900L..13A}, might contain a black hole with a mass within the pair-instability mass gap \cite{belczynski2016,spera2017,woosley2017,woosley2019,farmer2019,farmer2020,mapelli2020,farrell2020,tanikawa2020,renzo2020,vanson2020,costa2020}, suggesting that the system may have been formed by successive dynamical encounters in a dense environment, and that this channel of formation may represent a non negligible fraction of the mergers \cite{mapelli2016,rodriguez2019,dicarlo2019,dicarlo2020a,dicarlo2020b, fragione2020quad,fragione2020b,rizzuto2020,fishbach2020, gayathri2020,liulai2021, Rice2020, romero-shaw2020,palmese2020,
 safarzadeh2020,samsing2020,doctor2020,kimball2020a,mapelli2021,vigna2021}. Here, we study the contribution to the GW background of dynamical compact binaries formed in young star clusters  using the  simulated population presented in \cite{santoliquido2020a}. Young star clusters are a common birthplace of massive stars \cite{lada2003,portegieszwart2010} and a favorable environment for the dynamical assembly of BBHs \cite{banerjee2010,ziosi2014,mapelli2016,banerjee2017,banerjee2018,dicarlo2019,banerjee2020,kumamoto2019,rastello2019,kumamoto2020}. As described in \cite{dicarlo2020b,rastello2020}, we generated our dynamical binary catalogs with the direct $N$-body code \texttt{NBODY6++GPU} \cite{wang2015}, interfaced with the population-synthesis code \texttt{MOBSE} \cite{mapelli2017,giacobbo2018a}. For comparison, we also consider a population of isolated binaries obtained with the same population-synthesis code. In these simulations, the merger rate  takes into account the star formation history and the metallicity evolution in the Universe \cite{santoliquido2020a}.

In Section \ref{sec:catalogs}, we present the models used to compute  realistic catalogs of compact binaries; Section \ref{sec:spectra} describes the spectral properties of the GW background ; In Section~\ref{sec:background}, we calculate the GW background for the different models; Finally, in Section~\ref{sec:scenarios}, we discuss the possible detection scenarios.

\section{Catalog description}\label{sec:catalogs}

\subsection{Isolated CBCs}
The isolated binary mergers have been simulated with the population-synthesis code {\sc mobse} \cite{mapelli2017,giacobbo2018a,giacobbo2018b,giacobbo2020}. {\sc mobse} is a vigorous upgrade\footnote{{\sc mobse} is publicly available for download at \url{http://demoblack.com/catalog_codes/mobse-public-version/}.} of {\sc bse} \cite{hurley2000,hurley2002}, including up-to-date stellar wind models and recent prescriptions for electron-capture \cite{giacobbo2019}, core-collapse \cite{fryer2012} and pair instability supernovae \cite{mapelli2020}. In particular, we assume the rapid core-collapse supernova model by \cite{fryer2012} and we draw the natal kicks  from $v_k=(1-f_{\rm fb})\,{}v_{\rm H05}$, where $v_{\rm H05}$ is randomly drawn from a Maxwellian distribution function with one-dimensional root mean square $\sigma{}=15$ km s$^{-1}$ and $f_{\rm fb}$ is the fallback parameter described in \cite{fryer2012}. These prescriptions yield a minimum black hole mass of $\sim{}5$ M$_\odot$ and a maximum one of $\sim{}65$ M$_\odot$. However, only black holes with masses up to $\sim{}45$ M$_\odot$ merge within a Hubble time in isolated binaries, because of envelope loss during common envelope \cite{bouffanais2019}. Binary evolution is implemented as in \cite{hurley2002}. In particular, we describe the common envelope with the $\alpha{}$ formalism, assuming a value of $\alpha{}=5$.

For the results presented here, we have simulated a total of $1.2\times{}10^8$ binaries, evenly divided among 12 metallicities: $Z=0.0002,$ 0.0004, 0.0008, 0.0012, 0.0016, 0.002, 0.004, 0.006, 0.008, 0.012, 0.016 and 0.02. Note that $Z=0.02$ is approximately the solar metallicity, according to the historical definition \cite{anders1989}. The initial mass of the primary component is sampled from a Kroupa mass function \cite{kroupa2001}, while the mass of the secondary, the orbital period and the orbital eccentricity are drawn from the distributions presented in \cite{sana2012}.  We refer to \cite{giacobbo2018b} for further details.

\subsection{Dynamical CBCs}
Our catalogs of dynamically formed CBCs are the result of $106'000$ direct $N$-body simulations of young star clusters, previously described in \cite{dicarlo2020b} and \cite{rastello2020}. Young star clusters are young ($\leq{}100$ Myr) and dense (central density $\geq{}10^3$ stars pc$^{-3}$) stellar systems. Even if they are not as massive and as long-lived as globular clusters, they represent the most common channel of formation of massive stars \cite{lada2003,portegieszwart2010}. For this reason, a large fraction of black holes might be influenced by the dynamics of their parent star cluster before it gets disrupted and releases most of its stellar content in the galaxy field. In \cite{mapelli2016} and \cite{dicarlo2019}, we have shown that the dynamics of a young star cluster deeply affects the properties of BBHs: about half of the BBHs form by dynamical exchanges (hereafter, exchanged binaries) and even those that come from original binaries (i.e., binary stars that were already present in the initial conditions) suffer from dynamical encounters significantly. In the following, we refer to exchanged BBHs as \textit{Exch} and to original BBHs as \textit{Orig}.

As a result of dynamical evolution, the mass ratios and the total masses of dynamical CBCs are significantly different from those of isolated CBCs: dynamics produces more massive mergers and with more extreme mass ratios. We even find that $\sim{}1$\% of all BBH mergers from young star clusters contain a black hole with mass in the pair instability mass gap \cite{dicarlo2020a}. 

The dynamical CBCs considered in this work originate from star clusters with mass $M_{\rm SC}$ ranging from 300 to 30'000 M$_\odot$, randomly generated according to $dN/dM_{\rm SC}\propto{}M_{\rm SC}^{-2}$, consistent with the mass function of young star clusters \cite{lada2003}. They have been simulated for 100 Myr with the direct $N$-body code {\sc nbody6++gpu} \cite{wang2015}, interfaced with {\sc mobse} to guarantee that the stellar and binary evolution are implemented in the same way as in the isolated CBCs. All stars with mass $>5$ M$_\odot$ are initially members of binary systems. The masses, mass ratios and orbital properties of the original binary stars are drawn as in \cite{sana2012}, for consistency with observations and with the isolated CBCs. We simulated star clusters with three different metallicities $Z=0.0002$, 0.002 and 0.02. We refer to \cite{dicarlo2020b} and \cite{rastello2020} for further details.

\subsection{Cosmological evolution}\label{sec:cosmo}
Both isolated and dynamical CBCs are evolved across cosmic time with the semi-analytic code \texttt{Cosmo}$\mathcal{R}$\texttt{ate} \cite{santoliquido2020a,santoliquido2020b}. The basic idea of \texttt{Cosmo}$\mathcal{R}$\texttt{ate} is that the redshift evolution of CBCs depends on the star formation rate (SFR) density evolution, on the stellar metallicity evolution and on the delay time between formation and merger of the binary star. In particular, the merger rate density $\mathcal{R}(z)$ is calculated as 
\begin{equation}
\label{eq:mrd}
   \mathcal{R}(z) = \int_{z_{\rm max}}^{z}\psi(z')\,{}\frac{dt(z')}{dz'}\,{}\left[\int_{Z_{\rm min}}^{Z_{\rm max}}\eta(Z) \,{}\mathcal{F}(z',z, Z)\,{}dZ\right]\,{}dz',
\end{equation}
where 
$Z_{\rm min}$ and $Z_{\rm max}$ are the minimum and maximum metallicity, $\psi{}(z')$ is the cosmic SFR density at redshift $z'$, $\eta(Z)$ is the merger efficiency, namely the ratio between the total number $\mathcal{N}_{\text{TOT}}(Z)$ of compact binaries (formed from a coeval population) that merge within an Hubble time ($t_{{\rm H}_0}=13.6$ Gyr) and the total initial mass $M_\ast{}(Z)$ of the simulation with metallicity $Z$, and $dt(z')/dz'=\left\{H_0\,{}(1+z')\,{}\left[\Omega_{\rm M}\,{}(1+z')^3+\Omega_\Lambda\right]^{1/2}\right\}^{-1}$. For the cosmological parameters $H_0$, $\Omega_{\rm M}$ and $\Omega_\Lambda$ we use the values from \cite{ade2016}. Finally, $\mathcal{F}(z',z,Z)$ is the merger rate of compact binaries that form at redshift $z'$ from stars with metallicity $Z$ and merge at redshift $z$ from our simulations: 
\begin{equation}
   \mathcal{F}(z',z,Z) = \frac{1}{\mathcal{N}_{\rm TOT}(Z)}\frac{d\mathcal{N}(z',z,Z)}{dt(z)}\,{}p(z',Z),
\end{equation}
where 
\begin{equation}
 p(z', Z) = \frac{1}{\sqrt{2 \pi\,{}\sigma_{Z}^2}}\,{} \exp\left\{{-\,{} \frac{\left[\log{(Z/{\rm Z}_\odot)} - \mu(z')\right]^2}{2\,{}\sigma_{Z}^2}}\right\}. 
\end{equation}
is the distribution of the logarithms of stellar metallicities $\log{(Z/{\rm Z}_\odot)}$ at a given redshift, assumed to be a normal distribution with mean $\mu{}(z')$ and standard deviation $\sigma{}_{Z}=0.20$ \citep[see][for  details]{santoliquido2020a}. 
Operatively, we calculate the term $\frac{d\mathcal{N}(z',z,Z)}{dt(z)}$ from our catalogs of isolated and dynamical CBCs by assuming that
\begin{equation}
 \frac{d\mathcal{N}(z',z,Z)}{dt(z)}\approx{}\frac{\mathcal{N}(z',z,Z)}{\Delta{}t(z)},  
\end{equation}
 where $\Delta{}t(z)$ is the time-step of the numerical integration of eq.~$\ref{eq:mrd}$ in \texttt{Cosmo}$\mathcal{R}$\texttt{ate} and $\mathcal{N}(z',z,Z)$ is the number of binary compact objects that form at redshift $z'$, from stars with metallicity $Z$ and merge at redshift $z$, extracted from our catalogs of isolated and dynamical CBCs. 
The SFR density evolution is described with the fitting formula by \cite{madau2017}, while for the stellar metallicity evolution we adopt the fit by \cite{decia2018}, correcting it by the normalization from \cite{gallazzi2008}.

We calculate the merger rate density evolution separately for isolated and dynamical CBCs. In the dynamical (isolated) case, we assume that all the star formation happens in young star clusters (isolated binaries). Finally, we estimate the uncertainty on the merger rate density by varying the normalization of the SFR and the slope and the normalization of the metallicity evolution within one standard deviation, assuming that the observational uncertainties follow a Gaussian distribution. The local merger rates from \cite{santoliquido2020a} and used in the catalogue simulations are given in Table~\ref{MRD}. The optimistic and the pessimistic model that we will present in the results are obtained by considering the 50\% credible interval around the fiducial merger rate density evolution. 

\begin{table}[h]
\centering
\begin{tabular}{l|l|l|l|}
\cline{2-4}
                                & BBHs             & BNSs              & BHNSs            \\ \hline
\multicolumn{1}{|l|}{Dynamical} & 64$^{+34}_{-20}$ & 151$^{+59}_{-38}$ & 41$^{+33}_{-23}$ \\ \hline
\multicolumn{1}{|l|}{Isolated}  & 50$^{+71}_{-37}$ & 283$^{+97}_{-75}$ & 49$^{+48}_{-34}$ \\ \hline
\end{tabular}
\caption{Local merger rates in Gpc$^{-3}$yr$^{-1}$ from \texttt{Cosmo}$\mathcal{R}$\texttt{ate} \cite{santoliquido2020a} used in the catalogue simulations for both dynamical and isolated channels.}
\label{MRD}
\end{table}

\section{Spectral properties of the background}\label{sec:spectra}

The stochastic background is defined as the superposition of all sources that are not resolved by the detectors. It can be characterized at the observed frequency 
$f = f_s/(1+z)$, where $f_s$ the frequency in the source domain and $z$ the redshift, by the dimensionless quantity \cite{Allen_Romano_1999}
\begin{equation}
    \Omega_{gw}(f) = \frac{1}{\rho_c}\frac{d\rho_{gw}}{d\ln{(f)}}.
    \label{def}
\end{equation}
In the above formula, $\rho_c= \frac{3H_0^2c^2}{8\pi G}$ is the critical energy density of the Universe and $\rho_{gw}$ the gravitational energy density.

For compact binaries, the gravitational energy density is given by:
\begin{equation}
    \Omega_{gw}(f) = \frac{1}{c\rho_c}fF(f),
    \label{omg}
\end{equation}
In this expression, the total flux is the sum: 
\begin{equation}
    F(f) = T^{-1}\sum_{k=1}^{N}\frac{1}{4\pi r^2}\frac{dE_{gw}^k}{df}(f) , 
    \label{flux}
\end{equation}
where $N$ is the number of sources during the observation time $T$. The spectral energy density of any individual source $k$ is given by the relation : 
\begin{equation}
    \frac{1}{4\pi d_L^2}\frac{dE_{gw}^k}{df}(f) = \frac{\pi c^3}{2G}f^2 \tilde{h}_k^2(f),
    \label{Egw_circ_h}
\end{equation}
where
\begin{equation}
    \tilde{h}^2_{k}(f) = \tilde{h}^2_{+,k}(f)+\tilde{h}^2_{\times,k}(f),
\end{equation}
is the sum of the squared Fourier domain GW amplitudes of the two polarisations +/$\times$ given by:

\begin{align}
&\tilde{h}_{+,k}(f) = h_{z,k}\frac{1+\cos{}^2(\iota_k)}{2}\Gamma(f)\\  
&\tilde{h}_{\times,k}(f) = h_{z,k}\cos{(\iota_k)}\,{}\Gamma(f), 
\end{align}
with
\begin{equation}
    h_{z,k}=\sqrt{\frac{5}{24}}\,{}\frac{\left[G\mathcal{M}^{(z)}_k\right]^{5/6}}{\pi^{2/3}c^{3/2}d_L(z_k)}
    \label{h_z}
\end{equation}
where $\mathcal{M}^{(z)} = (m_1m_2)^{3/5} / (m_1+m_2)^{1/5}(1+z)$ is the corrected chirp mass, $z$ the redshift, $d_L$ the luminosity distance and $\iota$ the inclination angle of the binary. Finally, the function $\Gamma(f)$ encodes the evolution of the waveform as a function of the frequency in the different phases of the coalescence. 

In the case of BNSs and BHNS, we consider only the inspiral phase up to the last stable circular orbit $f_{LSO}=\frac{c^3}{6^{3/2}G \pi M}$ with $M=m_1+m_2$, which gives $\Gamma(f)=f^{-7/6}$. 

In the case of BBHs for which we consider the inspiral, merger and ringdown phases, the phenomenological waveforms of \cite{0710.2335}, calculated in the case of a circular orbit, give:
\begin{equation}
\Gamma(f= f_s/(1+z))=\left \{
\begin{array}{l l}
 (1+\sum^3_{i=2} \alpha _i\nu^i)f^{-7/6} & \text{if } f<f_{merg}  \\
 w_m (1+\sum^2_{i=1} \epsilon_i\nu^i)^2 f^{-2/3} & \text{if }f_{merg}\leq f<f_{ring}  \\
 w_r  \mathcal{L}^2(f,f_{ring},\sigma) & \text{if } f_{ring}\leq f<f_{cut}
 \end{array}
\right . \text{,}
\end{equation}
with 
\begin{equation}
\begin{array}{l}
\nu \equiv (\pi Mf)^{1/3},\\
\epsilon_1 = 1.4547\chi_{eff} - 1.8897,\\
\epsilon_2 = -1.8153\chi_{eff} + 1.6557,\\
\alpha_2 = -323/224 + 451\eta/168, \\
\alpha_3 = (27/8 -11\eta/6)\chi_{eff},
\end{array}
\end{equation}

$\mathcal{L}(f,f_{ring},\sigma)$
is the Lorentz function centered at $f_{ring}$ and with width $\sigma$, $w_m$ and $w_r$ are normalization constants ensuring the continuity between the three phases. 
In the expressions above, 
\begin{equation}
\eta=(m_1m_2)/M^2
\end{equation}
is the symmetric mass ratio and 
\begin{equation}
\chi_{eff} = \frac{(m_1\vec s_1 + m_2\vec s_2)}{M} \cdot{} \frac{\vec L}{L} 
\end{equation}
is the effective spin, a weighted combination of the projections of the individual spins $\vec s_1$ and $\vec s_2$ on the angular momentum $\vec L$. 

The frequencies at the end of the different phases, inspiral, merger and ringdown, and $\sigma$ ($\mu_k={f_1, f_2,\sigma,f_3}$) are calculated using Eq.~2 of \cite{0909.2867}:
\begin{equation}
\frac {\pi M}{c^3} \mu_k = \mu_k^0 + \sum_{i=1}^3 \sum_{j=0}^{N} x_k^{ij} \eta^i \chi_{eff}^j 
\end{equation}
where the coefficients $\mu_k^0$ and $x_k^{ij}$ are given in Table I of \cite{0909.2867}. 

Combining the expressions above, one obtains: 

\begin{equation}
    \frac{1}{4\pi d_L^2}\frac{dE_{gw}^{k,(C)}}{df}(f) =\frac{5 }{48 G}\,{}f^2 \frac{\left[G\mathcal{M}^{(z)}_k\right]^{5/3}}{\pi^{1/3}d_L^2(z)}\,{}\Gamma _k^2(f)\,{}F_\iota
    \label{waveform_circ}
\end{equation}
with $F_\iota  = \left[\frac{1+\cos{}^2(\iota)}{2}+\cos{(\iota)}\right]^2$ the inclination factor.

The eccentricity should not play a significant role for isolated binaries as demonstrated in \cite{2008.04890} but may have an impact if we consider dense environment like star clusters and should be taken into account. In this case, the spectral energy density of a binary with eccentricity $e_k$ is given for each harmonic $n$ by :
\begin{equation}
    \frac{dE_{gw}^{k,n}}{df}(f_n) = \frac{dE_{gw}^{k,(C)}}{df}(f_n)\,{}\frac{g(n,e_k)}{\Psi(e)}\,{}\left(\frac{4}{n^2}\right)^{1/3}
    \label{waveform_ecc}
\end{equation}
where $f_{n} =n\,{}f_{orb}/(1+z)$ is the observed frequency for the harmonic $n$. The case $n=2$ corresponds to the circular orbit.
The function $g(n,e)$ is a sum of Bessel functions:
 \begin{equation}
\begin{array}{c c}
 g(n,e)= \frac{n^4}{32} \left\lbrace\left[J_{n-2}(ne) - 2eJ_{n-1}(ne) + \frac{2}{n}J_n(ne) + 2eJ_{n+1}(ne)-J_{n+2}(ne)\right]^2 \right. \\ 
  \left. +(1-e^2)\left[J_{n-2}(ne) -2eJ_n(ne)+J_{n+2}(ne)\right]^2 + \frac{4}{3n^2}\left[J_n(ne)\right]^2\right\rbrace
\end{array}
\end{equation}
and
\begin{equation}
\Psi(e) = \frac{1+73/24e^2+37/96e^4}{(1-e^2)^{7/2}}
\end{equation}

Combining equations \ref{flux}, \ref{Egw_circ_h}, \ref{waveform_circ} and \ref{waveform_ecc}, we obtain the following expression for the total gravitational wave energy density: 
\begin{equation}
     \Omega_{gw}(f) = T^{-1}\frac{5}{18}\frac{\pi^{2/3}G^{5/3}}{H_0^2c^3}\sum_{k=1}^{N}\sum_n f_n^3 \frac{[\mathcal{M}^{(z)}_k]^{5/3}}{d_L(z_k)^2} \frac{g(n,e)}{\Psi(e)} \Gamma_k^2(f_n)F_\iota \mathrm{.}
\end{equation}

In this work, the energy density of a type of source $\Phi =\left\{ \mathrm{BBH, BNS, BHNS, All}\right\}$ is referred to as $\Omega_{gw}^{\Phi}(f)$ and calculated as the sum of the two populations of isolated (\textit{Iso}) and dynamical (\textit{Dyn}) sources.
\begin{equation}
    \Omega_{gw}^{\Phi}(f) =  (1-f_{Dyn}) \times\Omega_{gw}^{Iso,\Phi}+f_{Dyn}\times\Omega_{gw}^{Dyn,\Phi},
\end{equation}

where $\Omega_{gw}^{Iso,\Phi}$ and $\Omega_{gw}^{Dyn,\Phi}$ are derived from simulated catalogs of sources following the procedure described in section IV of \cite{2008.04890}, and where $f_{Dyn} = N_{Dyn} / (N_{Iso}+N_{Dyn})$ is the fraction of dynamical binaries.

The catalogs provide the parameters (masses, redshift, semi-major axes, eccentricity), for sources between redshifts $z\in[0,\,{}15]$ and for one year of observations.

Compared to \cite{2008.04890}, we have changed the spin distribution and have drawn the spin magnitudes $s_1$ and $s_2$ from a Maxwellian distribution with $\sigma=0.1$, in agreement with recent observations \cite{2010.14533}. For isolated sources, we assume the two spins to be aligned, while we consider a uniform distribution of the spin orientations for dynamical sources. 
We also assume a uniform distribution of the inclination angle.


\section{Total background}\label{sec:background}
In this Section, we present what we call the total background, i.e the sum of GW signals from all the sources all over the Universe, independently of the detectors.


\begin{figure}[h!]
\centering
  \includegraphics[width=18.5cm]{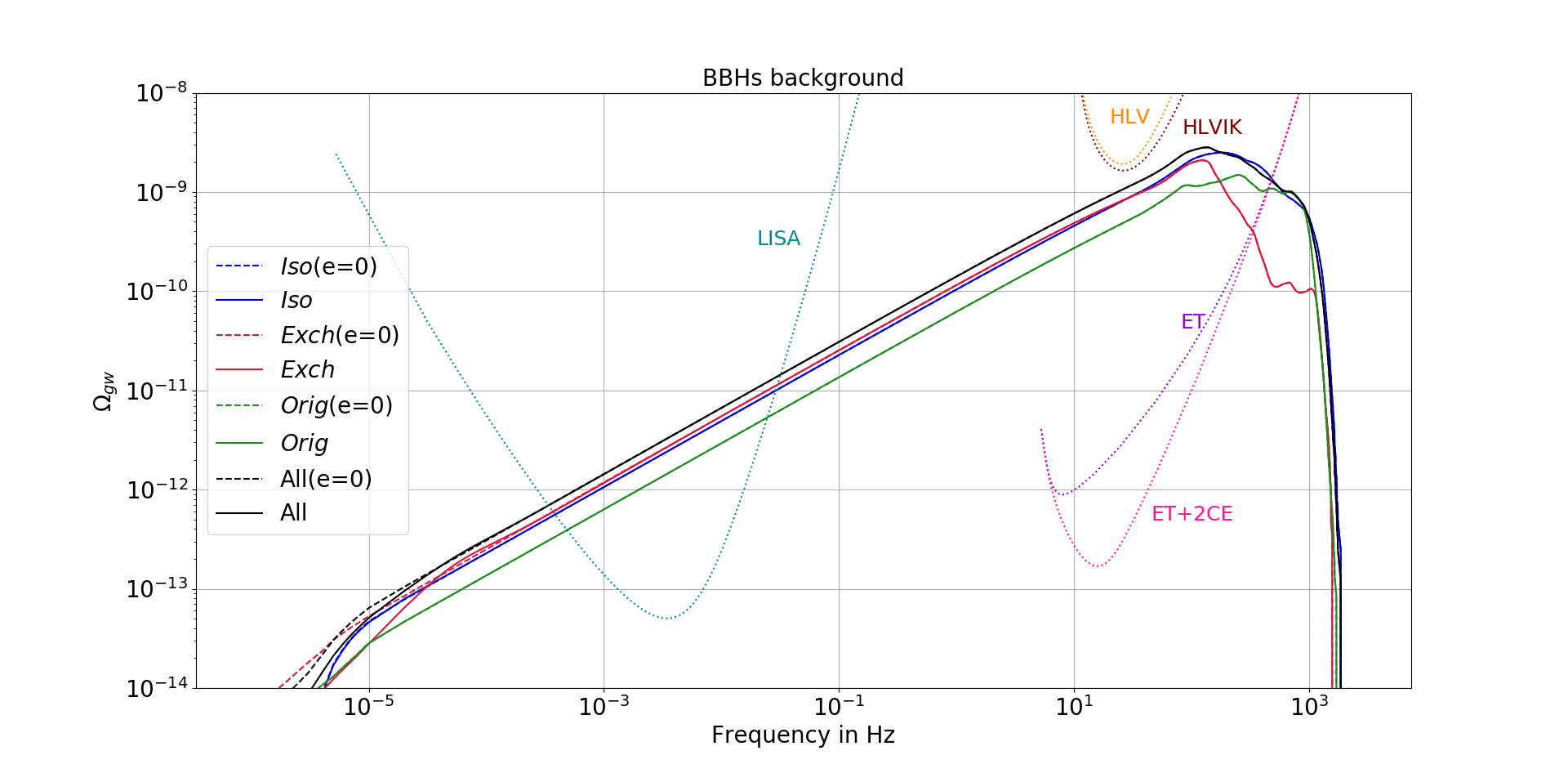}
\caption{Background energy density spectrum for BBHs by separating exchanged (red, \textit{Exch}), original (green, \textit{Orig}), isolated (blue, \textit{Iso}) and all binaries (black, \textit{Iso}+\textit{Orig}+\textit{Exch}) assuming $f_{Dyn}$=0.5.}
\label{BBH}
\end{figure}

\begin{figure}[h!]
\centering
  \includegraphics[width=18.5cm]{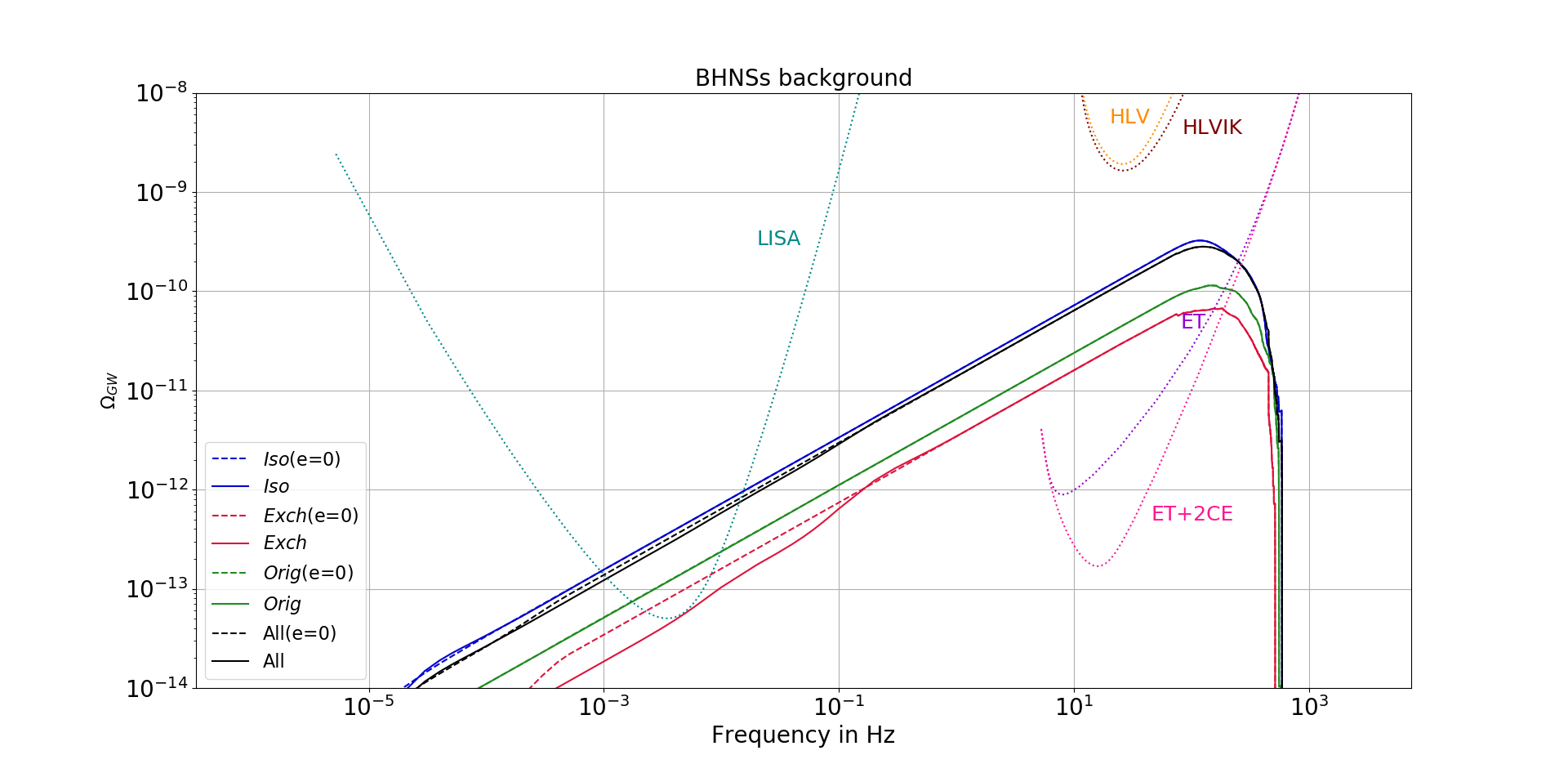}
\caption{Background energy density spectrum for BHNSs by separating exchanged (red, \textit{Exch}), original (green, \textit{Orig}), isolated (blue, \textit{Iso}) and all binaries (black, \textit{Iso}+\textit{Orig}+\textit{Exch}) assuming $f_{Dyn}$=0.5.}
\label{BHNS}
\end{figure}

\begin{figure}[h!]
\centering
  \includegraphics[width=18.5cm]{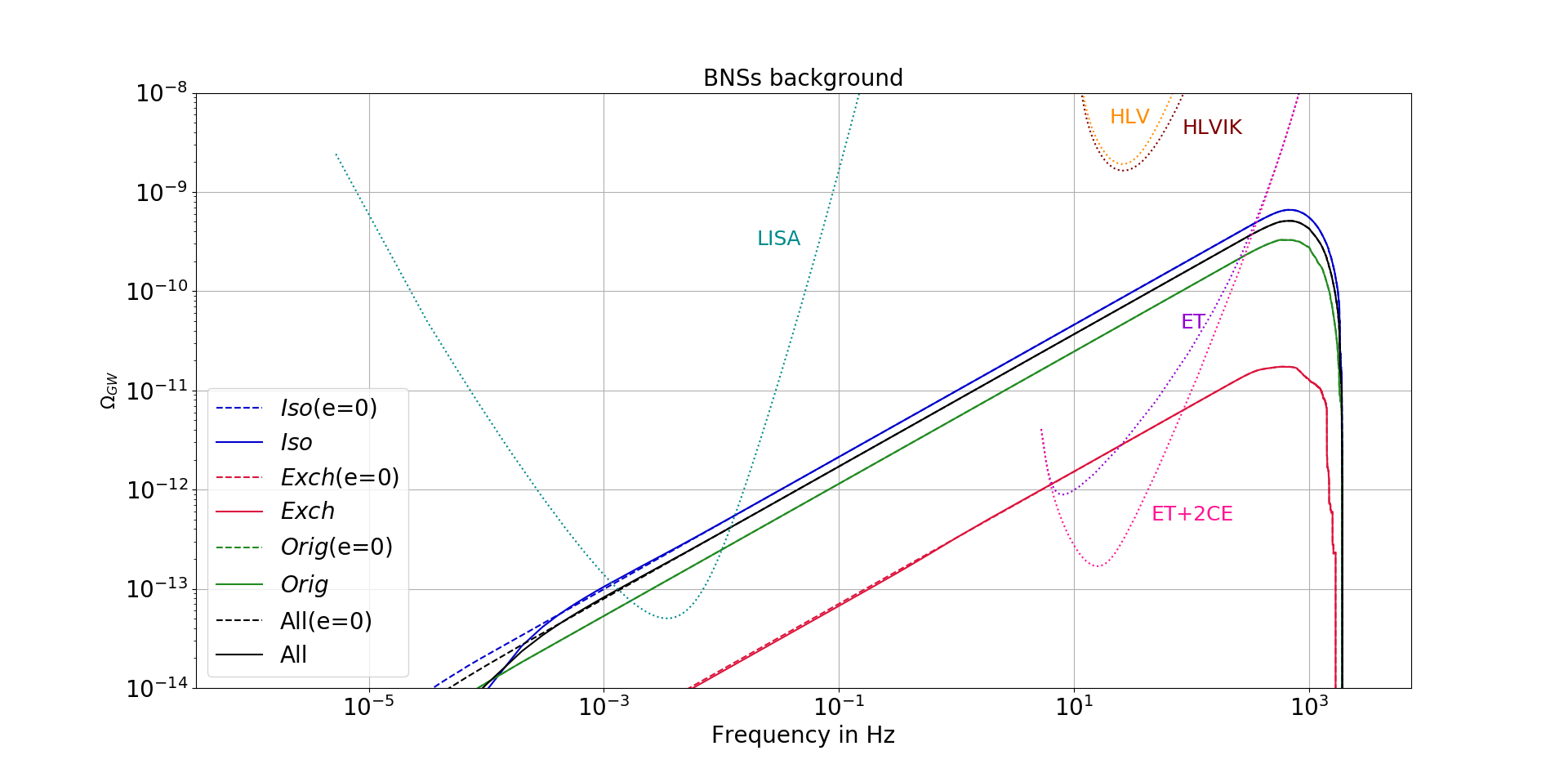}
\caption{Background energy density spectrum for BBHs by separating exchanged (red, \textit{Exch}), original (green, \textit{Orig}), isolated (blue, \textit{Iso}) and all binaries (black, \textit{Iso}+\textit{Orig}+\textit{Exch}) assuming $f_{Dyn}$=0.5.}
\label{BNS}
\end{figure}

\begin{figure}[h!]
\centering
  \includegraphics[width=18.5cm]{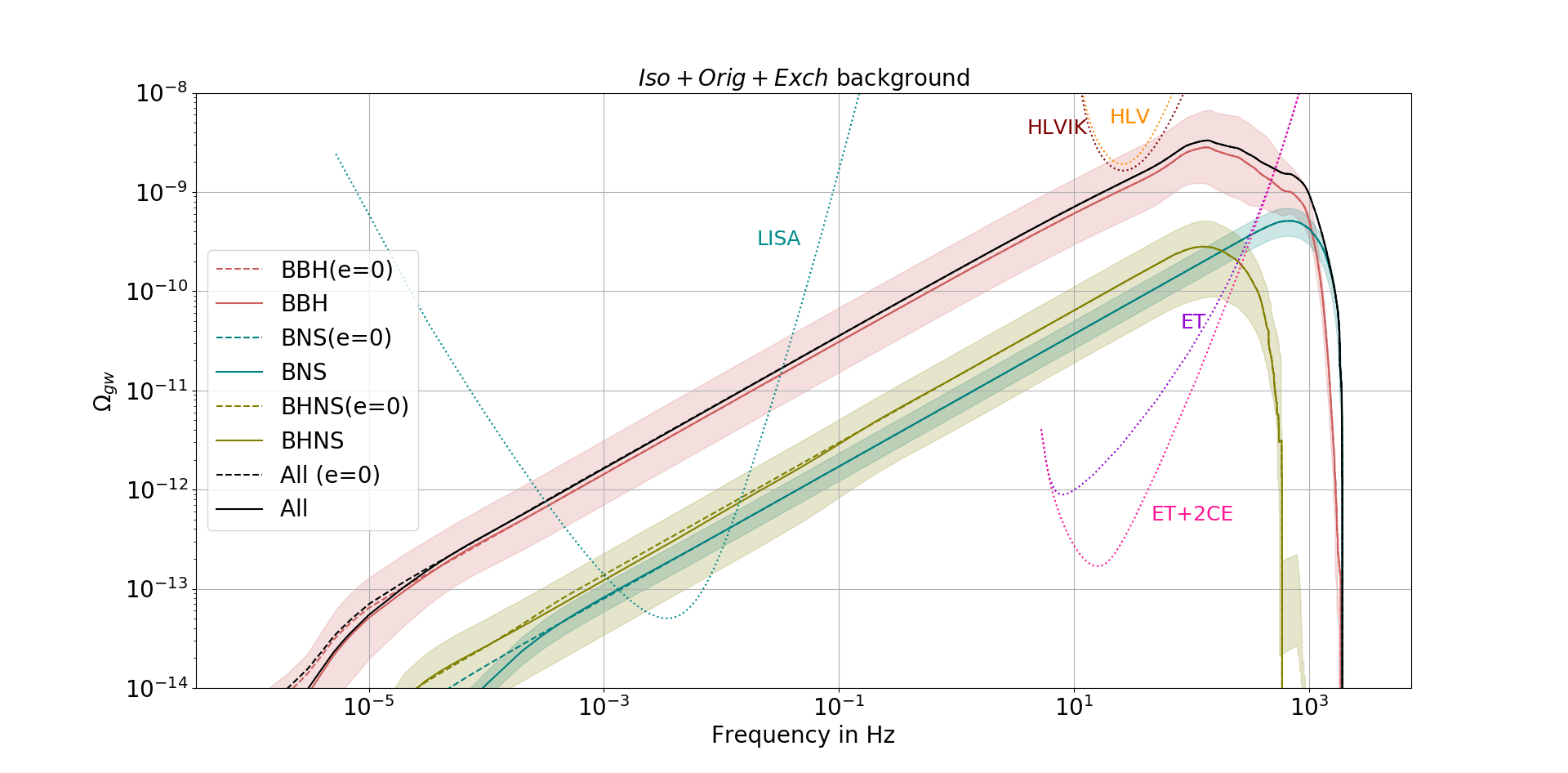}
\caption{Summary of the total energy density (\textit{Iso}+\textit{Orig}+\textit{Exch}) for each type of binaries: BBHs (red), BNSs (blue), BHNSs (green) and All (black, BBHs+BNSs+BHNSs) with $f_{Dyn}=0.5$.}
\label{tot}
\end{figure}

 In Fig.~\ref{BBH}, \ref{BHNS} and \ref{BNS}, we present  the contribution to the GW energy density $\Omega_{gw}(f)$ of the different formation channels, for each type of binaries (BBH, BNS, and BHNS). In particular, for the dynamical channel we distinguish between exchanged (\textit{Exch}) and original (\textit{Orig}) binaries. The relative fraction between exchanged and original binaries comes directly from the N-body simulations and depends on, e.g., the metallicity \cite{dicarlo2020b}. In order to highlight the impact of the eccentricity, we show the case of circular orbits ($e=0$) for comparison. Fig.~\ref{tot} shows the contribution  of the different types of binaries and their sum. Here,  we assume that dynamical binaries represent half of the population, i.e $f_{Dyn}=0.5$.


In all the plots, the dotted lines indicate the projected sensitivities, the so-called power integrated (PI) curves as defined in \cite{Thrane_2013}, for the space antenna LISA and for different terrestrial detector networks: 
\begin{itemize}
\item HLV: Advanced LIGO Hanford (H) and Livingston (L, \cite{aLIGO}), and Advanced Virgo (V, \cite{adVirgo}) at design sensitivity. 
\item HLVIK: HLV with in addition LIGO India (I, \cite{Indigo}), whose sensitivity will be similar to the two LIGO detectors, and the Japanese detector Kagra (K, \cite{Kagra}), also at design sensitivity.
\end{itemize}
For comparison, we also show the 3G designed power integrated curves of Einstein Telescope (ET\cite{ET}) and ET+2CE by adding two Cosmic Explorers (CE\cite{CE}).
A power-law stochastic background that is tangent to a PI curve is detectable with a signal-to-noise-ratio of 2. For LISA, we assume an effective integration time of 5 years (corresponding to the 10 years mission with a duty cycle of about 50\%) and for terrestrial detectors we assume an effective integration time of 1 year following \cite{1710.05837}.
The error bands shown in Figure~\ref{tot} (lower panel) represent the uncertainty of about a factor of two computed 
by combining the top and upper quartiles of the star formation rate $SFR(z)$ and the metallicity-redshift relation function described in the catalog section. 

For all types of binaries and formation channels, we can recognize the evolution as $\Omega_{gw}(f)\sim f^{2/3}$ coming from the inspiral stage, followed by a maximum and a sharp decrease. For BBHs, for which we include merger and ringdown, we observe a change of slope before the maximum due to the merger phase ($\Omega_{gw}(f)\sim f^{5/3}$ ). The cutoff corresponds to the frequency when all the sources have stopped emitting, which is around 1500~Hz for BBHs, 600~Hz for BHNSs and as high as 2000~Hz for BNSs because of their low masses. The background is thus dominated by BBHs below 1000 Hz, and then by BNSs, which remain the only sources at higher frequencies. 

The eccentricity does not play a significant role and the effects are only visible at the lowest frequencies, except for the case of exchanged BHNSs, where the difference with the circular case is noticeable up to about 0.1 Hz. However the background from BHNSs is low compared to the one from BBHs and the effect of the eccentricity does not appear in the total background. The values of $\Omega_{gw}$ at a reference frequency of 25 Hz are given in Table~\ref{tot_table}.


For isolated BBHs, we find an amplitude $\Omega_{gw}^{BBH,Iso} (25\,{}\mathrm{Hz})= 8.0 ^{+14.01}_{-6.31}\times10^{-10} $, which is very close to the \texttt{StarTrack} (ST) pop I/II predictions of $\Omega_{gw}^{ST, BBH}(25\,{}\mathrm{Hz})=7.3\times 10^{-10}$ \cite{2008.04890}. Our results are also in agreement with the recent predictions by the LIGO-Virgo-Kagra (LVK) collaboration of $\Omega_{gw}^{BBH,LVK}(25\,{}\mathrm{Hz}) = 5.0^{+1.7}_{-1.4}\times 10^{-10}$ \cite{UpperLimits_O3a}. The LVK error bars correspond to the standard Poisson uncertainty on the local merger rate and on the uncertainty on the parameters of the mass distribution, assumed to be a broken power law derived from 50 events observed in the second GW transient catalog \cite{GWTC2}. 

For isolated BNSs, we obtain $\Omega_{gw}^{BNS,Iso}(25\,{}\mathrm{Hz}) = 8.5_{-2.49}^{+3.35}\times10^{-11}$, which is within the error bars of the LVK prediction of $\Omega_{gw}^{BNS,LVK}(\mathrm{25Hz}) = 2.1^{+2.9}_{-1.6} \times 10^{-10}$, calculated assuming a local rate derived from the observation of two events and a uniform distribution of the component masses between $1-2.5$. On the other hand, our prediction is a factor $\sim$ 8 higher than the one derived from ST ($\Omega_{gw}^{BNS,ST}(25\,{}\mathrm{Hz}) =1.0 \times 10^{-11}$, \cite{2008.04890}). 

Regarding BHNSs, we find an amplitude of $\Omega_{gw}^{BHNS,Iso}(25\,{}\mathrm{Hz}) = 1.3^{+1.38}_{-1.01} \times 10^{-10}$, which is in agreement with the LVK upper limit \cite{UpperLimits_O3a} $\Omega_{gw}^{BHNS,LVK}(25\,{}\mathrm{Hz})<8.4\times 10^{-10}$, but one order of magnitude larger than the ST pop.~I/II prediction of $\Omega_{gw}^{BHNS,ST}(25\,{}\mathrm{Hz}) = 1.4\times 10^{-11}$  \cite{2008.04890}. This difference with respect to ST is expected, because our isolated BNS and BHNS merger rate density (see \cite{santoliquido2020a}) is about one order of magnitude higher than the one obtained from the ST simulations adopted in \cite{2008.04890}. The main reason for this is the different natal kick prescription.

Including the dynamical population, and assuming $f_{Dyn}=0.5$, the background increases by a factor of 1.6 for BBHs, and sligthly decreases for BHNSs and BNSs by factors of 1.3 and 1.2, respectively.

Adding together the isolated and dynamical populations and all types of binaries (BNS + BBH + BHNS), we find  $\Omega_{gw}^{All,All}(25\,{}\mathrm{Hz}) = 1.2^{+1.38}_{-0.65} \times 10^{-9}$,  which is below the most stringent upper limit of $\Omega^{UL}_{gw}(25\,{}\mathrm{Hz}) = 3.4 \times 10^{-9}$ (log-uniform prior), derived from the data of the three first science runs of LVK \cite{UpperLimits_O3a}, for $\Omega_{gw} \sim f^{2/3}$.

\begin{table}[]
\centering
\begin{tabular}{|l|l|lll|l|}
\hline
     & \textit{Iso}     & \textit{Orig} & \textit{Exch} & \textit{Dyn}(\textit{Orig}+\textit{Exch}) & All ($f_{Dyn}$=0.5)     \\ \hline
BBH  & 8.0$\times$10$^{-10}$ & 4.7$\times$10$^{-10}$  & 8.1$\times$10$^{-10}$ & 1.3$\times$10$^{-9}$ & 1.0$\times$10$^{-9}$  \\
BNS  & 8.5$\times$10$^{-11}$ & 4.6$\times$10$^{-11}$  & 2.8$\times$10$^{-12}$ & 4.8$\times$10$^{-11}$ & 6.7$\times$10$^{-11}$  \\
BHNS & 1.3$\times$10$^{-10}$ & 4.4$\times$10$^{-11}$  & 2.9$\times$10$^{-11}$ & 7.3$\times$10$^{-11}$ & 1.0$\times$10$^{-10}$  \\ \hline
All  & 1.0$\times$10$^{-9}$ & 5.6$\times$10$^{-10}$& 8.4$\times$10$^{-10}$ & 1.4$\times$10$^{-9}$  & 1.2$\times$10$^{-9}$  \\ \hline
\end{tabular}
\caption{Energy density ($\Omega_{gw}$ at 25 Hz) of the total background for isolated, dynamical exchanged and original binaries (separately and the sum), and the sum of isolated and dynamical binaries, assuming an equal fraction of each.}
\label{tot_table}
\end{table}


\section{The background from BBHs}\label{sec:scenarios}

As seen in the previous Section, the background from BBHs dominates in the frequency band where the detectors are the most sensitive. Here, we study in more details the impact of the different parameters and the model uncertainties on the BBHs energy density spectrum.

\paragraph{Impact of the mass distribution} 
For BBHs, the contributions from $Exch$ and $Orig$ exhibit some bumps above $80$ Hz (see Fig.~\ref{BBH}) which are related to the mass and redshift distributions. We investigate this effect by plotting  the contributions from different redshifted chirp mass ranges separately.

\begin{figure}[h!]
\centering
  \includegraphics[width=17cm]{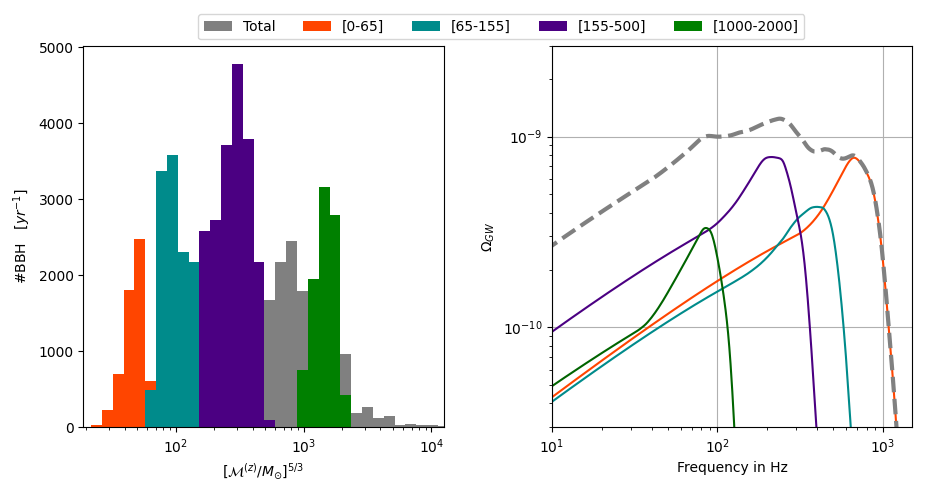}
\caption{Sub-populations 
of the redshifted chirp mass for original (\textit{Orig}) BBHs: $[0-65]$ in orange, $[65-155]$ in blue, $[155-500]$ in purple, $[1000-2000]$ in green and the total in grey (the ranges are given in units of ${\rm M}_\odot^{5/3}$). Left: truncated histograms of the redshifted chirp mass for the 4 sub-populations and the total. Right: Corresponding energy density spectra.}
\label{BBHori}
\end{figure}

\begin{figure}[h!]
\centering
  \includegraphics[width=17cm]{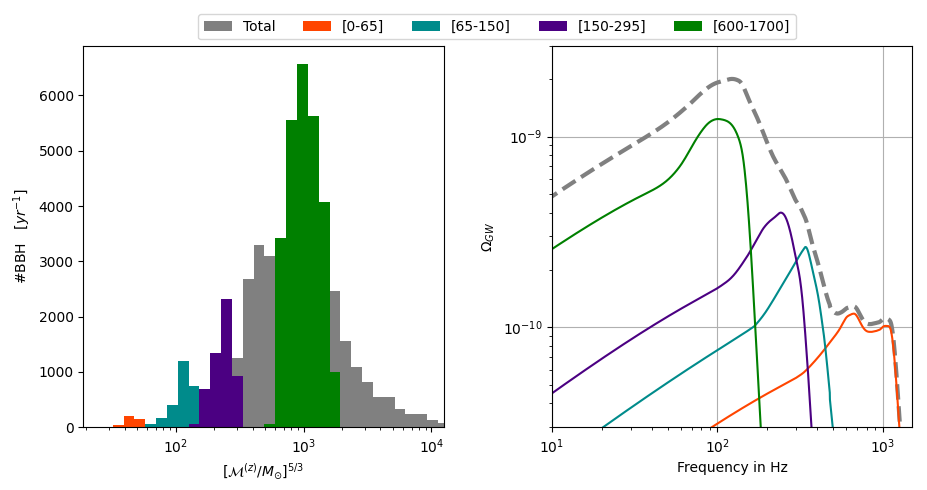}
\caption{Sub-populations 
of the redshifted chirp mass for exchanged (\textit{Exch}) BBHs: $[0-65]$ in orange, $[65-155]$ in blue, $[155-295]$ in purple, $[600-1700]$ in green and the total in grey (the ranges are given in units of ${\rm M}_\odot^{5/3}$). Left: histograms of the redshifted chirp mass for the 3 sub-populations and the total. Right: Corresponding energy density spectra.}
\label{BBHexc}
\end{figure}

In Figs.~\ref{BBHori} and \ref{BBHexc}, the left-hand panels show the histograms of the redshifted chirp mass for original and exchanged BBHs, with different colors for three different ranges and for the total distribution. The right-hand panels show the corresponding spectra.   
We highlight the strong impact of the redshifted chirp mass on the shape of the spectrum. For example, in Fig.~\ref{BBHori} the orange sub-population ($[\mathcal{M}_c(1+z_m)]^{5/3}$ in the range 0--65 M$_{\odot}^{5/3}$) is clearly responsible for the last bump at 660 Hz.
In the case of exchanged binaries in Fig.~\ref{BBHexc}, separating the contributions from the different redshifted chirp mass ranges does not allow us to separate the two high frequency bumps at 680 Hz and 1060 Hz (orange sub-population). This irregular shape springs mainly from the convolution of the distribution of chirp masses  with the distribution of  redshift. 
We discuss this in detail in Appendix~\ref{sec:A1}.  

\begin{figure}[h!]
\centering
  \includegraphics[width=17.5cm]{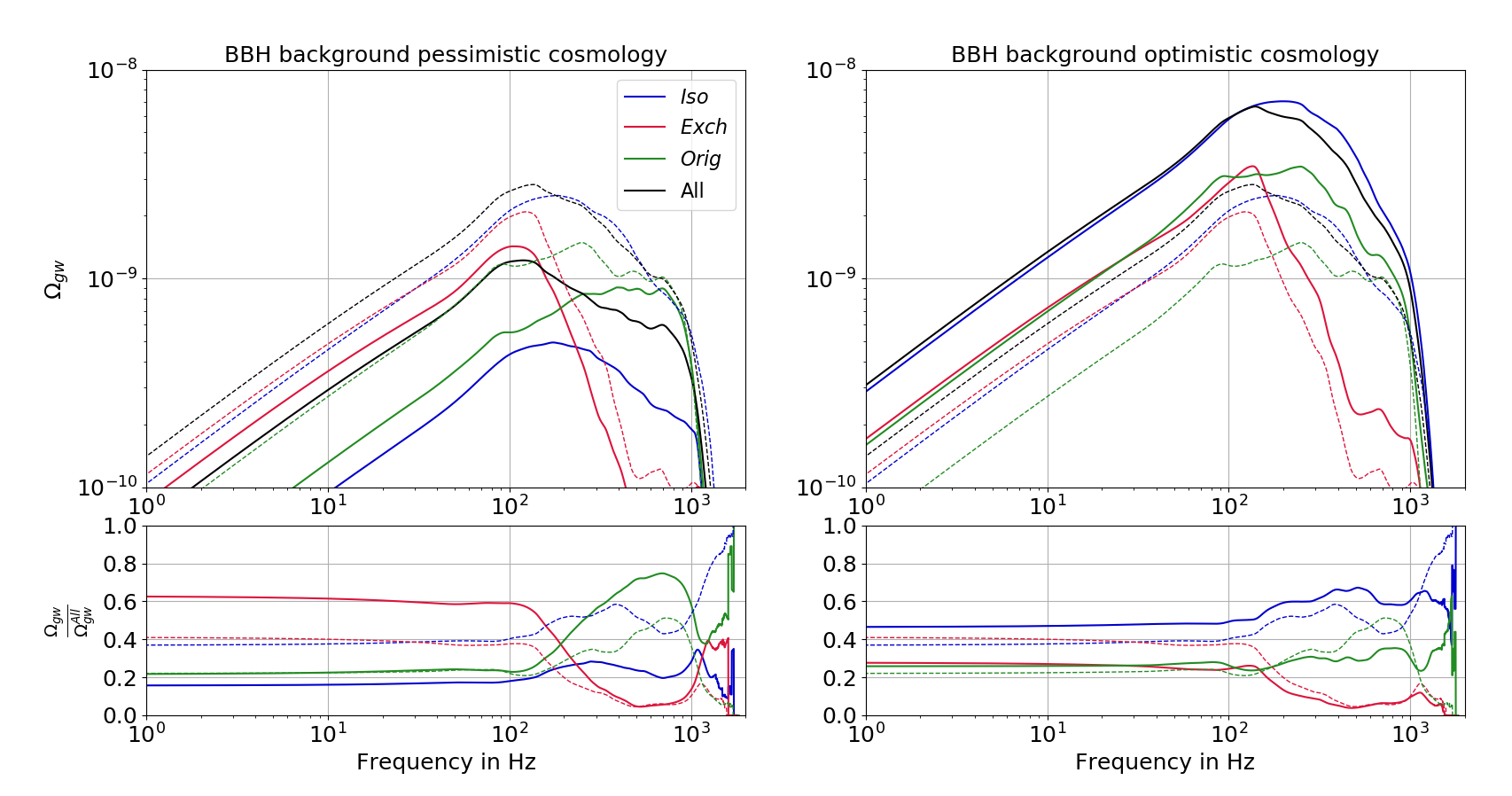}
\caption{Total BBH background (top). The total BBHs energy density is shown in black (assuming $f_{Dyn}$=0.5). The blue, red and green curves represent the energy density for isolated, dynamical exchanged and dynamical original binaries, respectively. The left (right) panel shows the pessimistic (optimistic) catalog. For the two cases, we have plot the ratio $\Omega^{subtype}_{gw} /\Omega^{All}_{gw} $ in the bottom panel, representing a predominance diagram between the different sub-types of BBHs. In all panels the dashed lines indicate the fiducial model, for comparison.}
\label{BBHerr}
\end{figure}

\paragraph{Impact of star formation and metallicity}
The shape of the background from BBHs reflects the distribution of the redshifted masses, and this distribution strongly depends on both the relations redshift-metallicity and redshift-star formation rate \cite{madau2017}. 
The uncertainties on the star formation model result in variations of a factor two of the spectrum amplitude (see Fig.~\ref{BBH}). For this calculation, we use the lower and upper quartiles 
of the merger rate density obtained with the code \texttt{Cosmo}$\mathcal{R}$\texttt{ate} by changing both the star formation rate density and the metallicity slope and normalization within $1\,{}\sigma{}$  \cite{santoliquido2020a,santoliquido2020b}. This allows us to build a pessimistic and an optimistic catalog (see Section \ref{sec:cosmo}) from which we derive the corresponding energy density spectra. The top left (right) panel of Fig.~\ref{BBHerr} 
shows the energy density for the pessimistic (optimistic) model in solid lines. For comparison, the dashed lines indicate the fiducial model. The lower panel shows the proportion of every type of BBHs (\textit{Iso}, \textit{Exch}, \textit{Orig}) in terms of contribution to the BBHs energy density.

The spectrum for the first quartile on the left panel of Fig.~\ref{BBHerr} is dominated by the dynamical population (\textit{Exch} below 200 Hz and \textit{Orig} above). The uncertainties affect more the isolated population, yielding to a reduction of $\Omega_{gw}^{BBH, Iso}$ by a factor 5, while it is a factor 2 for original binaries and 1.5 for exchanged binaries. 

For the upper quartile, on the opposite, the dynamical population is buried below the isolated population that strongly dominates the background shape. This comes from the large uncertainties affecting the isolated population, giving $\Omega_{gw}^{Iso}(25\,{}\mathrm{Hz}) = 1.4\times 10^{-9}$ for the upper quartile, compared to $\Omega_{gw}^{Orig}(25\,{}\mathrm{Hz}) = 7.3\times 10^{-10}$ and $\Omega_{gw}^{Exch}(25\,{}\mathrm{Hz}) = 4.0\times 10^{-10}$ for dynamical original and exchanged binaries, respectively. The main reason for this large difference is that dynamical binaries are less affected by the metallicity of the progenitor stars than the isolated binaries, as already discussed by \cite{santoliquido2020a}. In the pessimistic case, metal-rich stars are more common than in the optimistic case, leading to a much lower merger rate of isolated BBHs.


\paragraph{Implication of dynamical BBH rate proportions on the CBC backgrounds}

\begin{figure}[h!]
\centering
  \includegraphics[width=17.5cm]{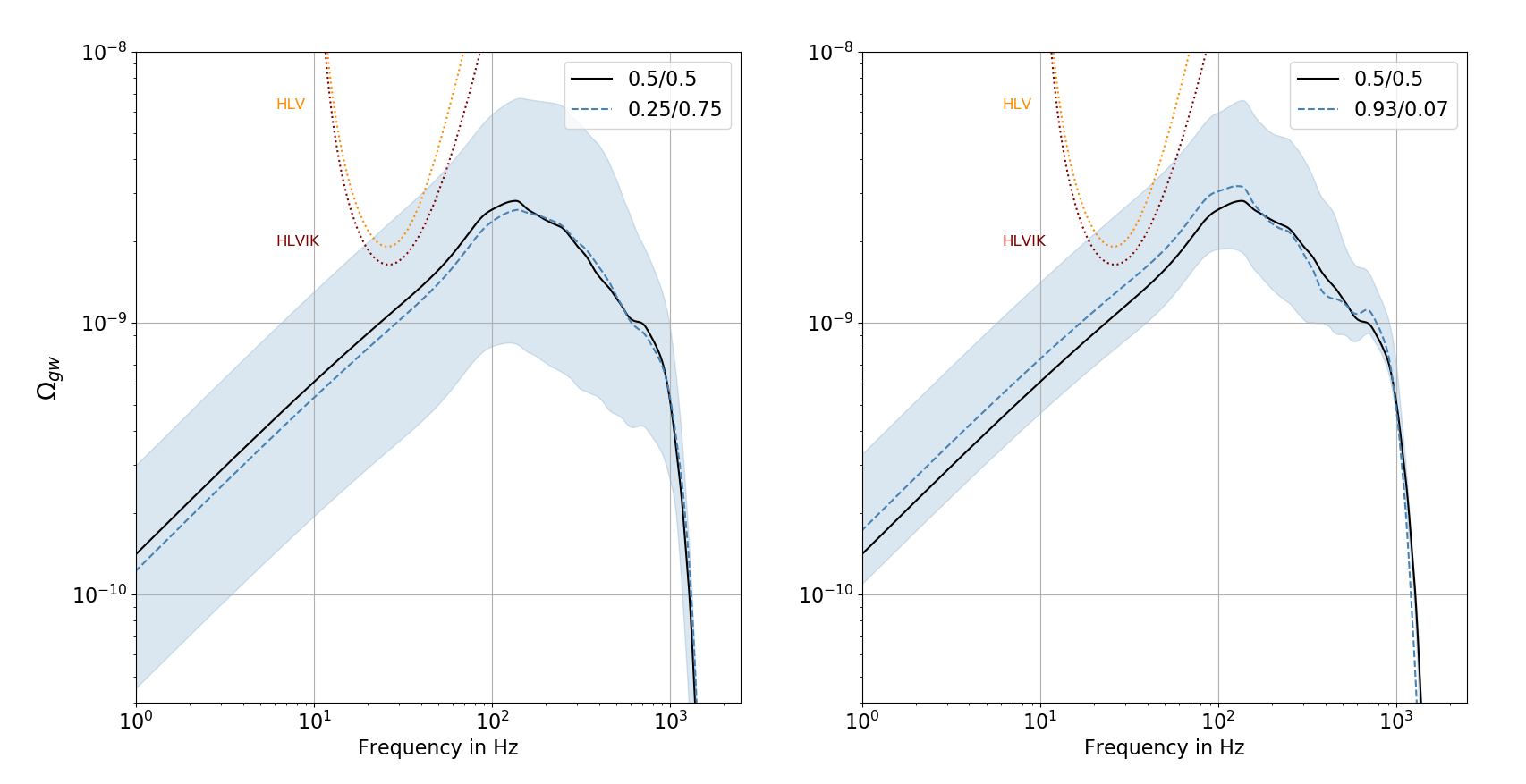}
\caption{Energy density for the population of BBHs considering the two extreme fractions ($f_{Dyn}/f_{Iso}$) of dynamical binaries inferred from the LIGO--Virgo catalog \cite{2010.14533} (Left: $f_{Dyn}=0.25$, right: $f_{Dyn}=0.93$), compared with the assumption of $f_{Dyn}=0.5$ (black line). The blue filled area corresponds to the uncertainties on star formation and metallicity, as discussed in the text.}
\label{BBH_prop}
\end{figure}

As seen previously, the dynamical population can play a significant role in the energy density spectrum from BBHs, but we still know little about the fraction of the population they represent. From the second LIGO--Virgo catalog \cite{2010.14533}, one can infer a fraction of dynamical binaries $f_{Dyn}$ between $0.25-0.93$ assuming $\chi_{eff}$ as an indicator of dynamical formation \footnote{This assume a systematic positive $\chi_{eff}$ for isolated binaries instead dynamical ones would lead to an asymmetric distribution around 0.}. Fig.~\ref{BBH_prop} shows the two extreme scenarios with $f_{Dyn} = 0.25$ (left-hand panel) and $f_{Dyn} =0.93$ (right-hand panel). The error band corresponds to the star formation/metallicity uncertainty detailed in the previous paragraph.

 The resulting values of $\Omega_{gw}$ at 25 Hz for the different scenarios are shown in Table~\ref{table_prop}.  The proportion between the two formation channels does not have a huge impact on the background amplitude. Considering the sole proportion uncertainty, we have a deviation of only several percents $\Omega_{gw}(25\,{}{\rm Hz}) = 1.0^{+0.21}_{-0.12} \times 10^{-9}$. However, the shape changes slightly depending on $f_{Dyn}$, in particular around the bump (at 100 Hz).
 
As already demonstrated, the uncertainties due to star formation and metallicity evolution are larger for the isolated population than for the dynamical one, and thus larger for a smaller dynamical fraction $f_{Dyn}$. 

\begin{table}[]
\centering
\begin{tabular}{|c|c|c|c|}
\hline
$f_{Dyn}$/$f_{Iso}$ & 0.25/0.75                             & 0.5/0.5                               & 0.93/0.07                             \\
$\Omega_{gw}$(25Hz)   & 0.9$^{+1.33}_{-0.59}\times$10$^{-9}$ & 1.0$^{+1.26}_{-0.54}\times$10$^{-9}$ & 1.3$^{+1.15}_{-0.47}\times$10$^{-9}$ \\ \hline
\end{tabular}
\caption{Energy density ($\Omega_{gw}$ at 25 Hz) of the sum of isolated and dynamical binaries assuming an equal proportion of the two populations, and for the two extreme fractions of dynamical binaries ($f_{Dyn}=0.25$ and $0.93$) inferred from the LIGO--Virgo catalog \cite{2010.14533}.}
\label{table_prop}
\end{table}

\section{Residual backgrounds and detectability}

The residual background, as opposed to the total background, is the sum of all the sources that cannot be resolved, either because they overlap or because they are too faint to be detected. In this Section,  estimate the residual background for two 2G detector networks, HLV and HLVIK, and calculate its detectability.

\subsection{Residual backgrounds for 2G detectors}
\begin{figure}[h!]
\centering
  \includegraphics[width=17.5cm]{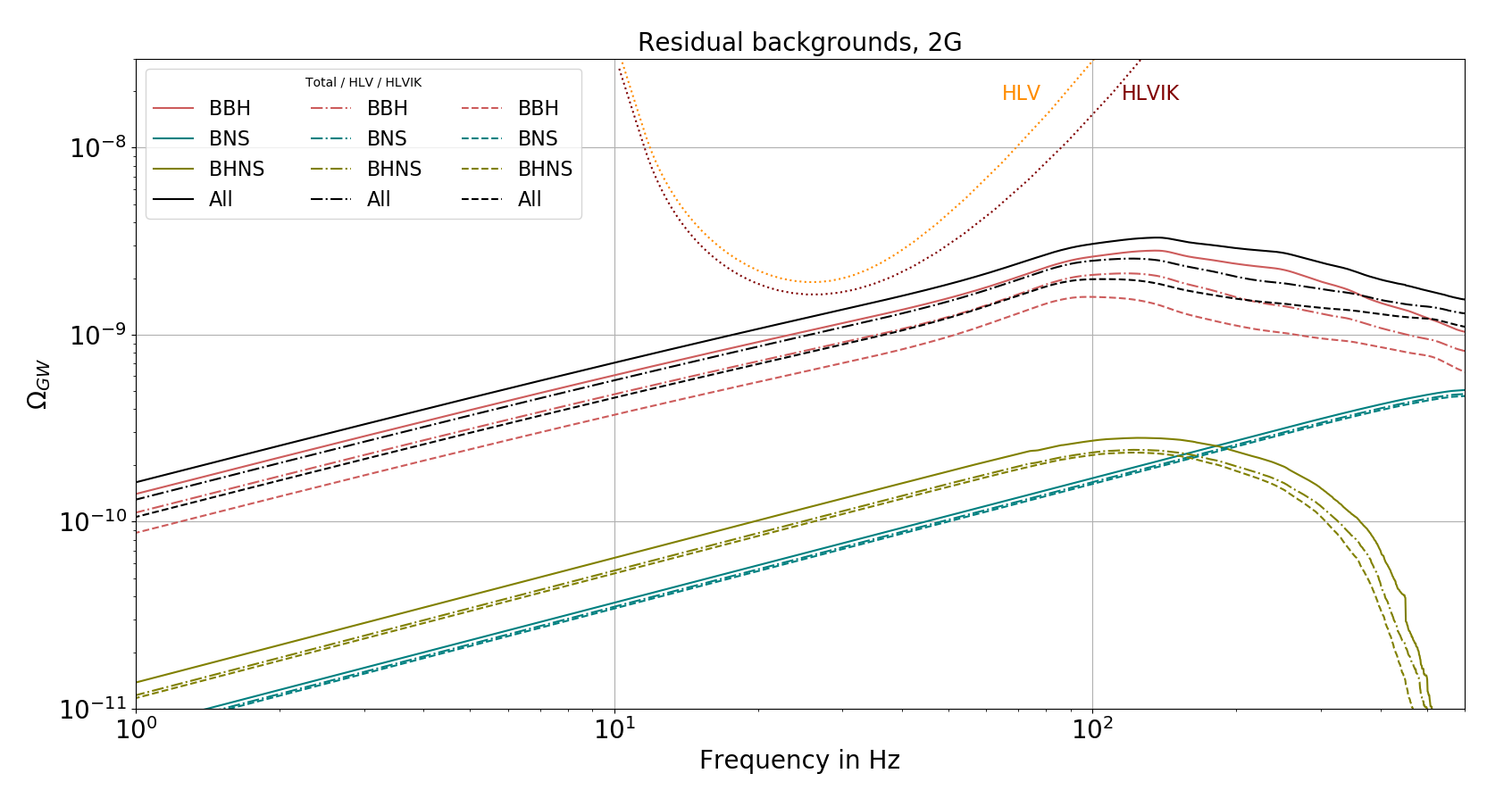} 
\caption{Residual background for 2G detector networks HLV(dash-dotted line) and HLVIK (dashed line). The total background is indicated for comparison (solid line). The different network sensitivities are represented by dotted lines (see previous Section).}
\label{2G}
\end{figure}

 Following the procedure described in section IV of \cite{2008.04890},
 we compute the residual background by removing all detected sources from the total population. We consider a source as detected when its signal-to-noise ratio $\rho$ (eq.~26 from \cite{2008.04890}) is larger than a threshold $\rho_{T}=12$. The overlap of detectable sources is very unlikely for 2G detectors and is ignored in this study. Fig.~\ref{2G} shows the residual background for HLV (dash-dotted line) and HLVIK (dashed line). The solid lines represent the total background. For this study we consider an equal fraction of isolated and dynamical binaries, i.e $f_{Dyn} = f_{Iso} = 0.5$.
 As the sensitivity of the detectors will improve in the future, they will be able to detect more sources, which will decrease the level of the residual background, assuming one can successfully subtract individual signals from the data \cite{1611.08943}. 
 
 
 Table~\ref{Res_tab} compares the value of $\Omega_{gw}$ at the reference frequency of 25~Hz for the total and the residual backgrounds, as well as the number of detected sources considering one year of observation. 
 
 In order to quantify the reduction of the background, we calculate the ratio $r_{\Omega}$ between the energy densities of the residual background and the total background at the most sensitive frequency $f_{ref}=25$~Hz:
\begin{equation}
r_{\Omega}=\frac{\Omega_{GW,res}(f_{ref})}{\Omega_{GW,tot}(f_{ref})}.\qquad 
\label{ratio}
\end{equation}
For comparison, we also calculate the ratio between the number of sources contributing to the residual background and the total number of sources:
\begin{equation}
 r_{count}=\frac{N_{res}}{N_{tot}}.
\label{ratio_Count}
\end{equation}

The ratios $r_{\Omega}$ and $r_{count}$ are also reported in percentages in Table~\ref{Res_tab}. 

\begin{table}[]
\begin{tabular}{|c|c|c|c|c||c|}
\hline
\multicolumn{2}{|c|}{}      & BBHs                  & BNSs                          & BHNSs                         & All                       \\ 
\hline
\multirow{2}{*}{Total}   & $\Omega_{gw}$ & 1.0$\times$10$^{-9}$ & 6.7$\times$10$^{-11}$ & 1.0$\times$10$^{-10}$ & 1.2$\times$10$^{-9}$ \\
                       & N$_{tot}$          &  106136             & 275337               & 151525               & 532898              \\
\hline
\multirow{2}{*}{HLV}   & $\Omega_{gw}$($r_\Omega$) & 8.2$\times$10$^{-10}$(78\%) & 6.5$\times$10$^{-11}$(97\%) & 1.0$\times$10$^{-10}$(97\%) & 9.9$\times$10$^{-10}$(80\%) \\
                       & \#N$_{det}$(1-$r_{count}$)          & 617 ($<$1\%)              & 5 ($\sim$0\%)               & 6 ($\sim$0\%)               & 628 ($\sim$0\%)              \\ \hline
\multirow{2}{*}{HLVIK} & $\Omega_{gw}$($r_\Omega$) & 6.4$\times$10$^{-10}$(61\%) & 6.3$\times$10$^{-11}$(95\%) & 9.7$\times$10$^{-11}$(94\%) & 8.0$\times$10$^{-10}$(65\%) \\
                       & \#N$_{det}$(1-$r_{count}$)          & 3051 ($\sim$3\%)                & 20 ($\sim$0\%)               & 37 ($\sim$0\%)               & 3108($<$1\%)               \\ \hline
\end{tabular}
\caption{Energy density $\Omega_{GW}$ at the reference frequency of 25 Hz, and number of detected sources $N_{det}$, with HLV and HLVIK, assuming $f_{Dyn} = f_{Iso} = 0.5$ and one year of observation. Within the round brackets, we show the ratio $r_\Omega$ and $1$-$r_{count}$ as defined in the text.}
\label{Res_tab}
\end{table}

As expected, the detected sources are the closest and the loudest and then the ones that give the largest contribution to $\Omega_{GW}$. For instance, only 0.6\% of BBHs are detected with HLV, but their subtraction reduces the energy density by 22\%. For BNSs and BHNSs number of detection rate is about  5--6 per year and it corresponds to a reduction of 3\% of $\Omega_{GW}$. When adding LIGO--India and Kagra in the network, the fraction of detected sources increases to 3\% for BBHs, which corresponds to a reduction of 39\% of $\Omega_{GW}$ at 25 Hz. For BNSs and BHNSs about 20 and 37 sources are detected, leading to a reduction of 5\% and 6\% of $\Omega_{GW}$. The fraction of detected sources and the reduction of the energy density are not impacted by the formation channels (\textit{Iso} and \textit{Dyn}), as shown in Appendix \ref{App_Prop}. Finally, combining BBHs, BHNSs and BNSs, we find $\Omega_{GW,res}$ at 25 Hz of 9.9$\times 10^{-10}$ for HLV and 8.0$\times 10^{-10}$ for HLVIK.

 \subsection{Detectability}
 
We now evaluate the detectability of the residual background, and we study a realistic detection scenario for the near future.
 
The optimal strategy to search for a stochastic background which can be confounded with the noise of a single detector is to cross correlate two (or several) detectors, in order to eliminate the noise and recover the common signal. For a pair of detectors $i$ and $j$, the signal-to-noise ratio (SNR) of the cross correlation statistic \cite{Allen:1997ad}, assuming independent uncorrelated noise in each detector, is given by
\begin{equation}
\text{SNR}_{ij} =\frac{3 H_0^2}{10 \pi^2}\,{} \sqrt{2T} \left[
\int_0^\infty df
\frac{\gamma_{ij}^2(f)\,{}\Omega_{gw}^2(f)}{f^6\,{} P_i(f)\,{}P_j(f)} \right]^{1/2}\,,
\label{eq:snrCC}
\end{equation}

where $\gamma_{ij}(f)$ is the normalized isotropic overlap reduction function (ORF), used to account for the reduction in sensitivity due to the separation and relative orientation of the two detectors \cite{1993PhRvD..48.2389F,1992PhRvD..46.5250C}. $P_i(f)$ and $P_j(f)$ are the one sided power spectral noise densities for detectors $i$ and $j$, and $T$ is the effective observation time. 

Combining different detector pairs for a $n$-detectors network \cite{1993PhRvD..48.2389F,1992PhRvD..46.5250C} we obtain: 
\begin{equation}
    SNR = \left[ \sum_{i=1}^n \sum_{j>i} SNR_{ij}^2\right]^{1/2}
\end{equation}
As an example, for the HLV case we can write
\begin{equation}
    SNR_{HLV} = [SNR_{HL}^2+SNR_{HV}^2+SNR_{LV}^2]^{1/2}
\end{equation}

The SNR for the two networks HLV and HLVIK at design sensitivity and for one year of observations are given in Table~\ref{SNR_Des}. For comparison, we provide the results for both the total and the residual backgrounds. 

\begin{table}
\centering
\begin{tabular}{|c|cc|}
\hline
  & HLV        & HLVIK    \\ \hline
Total  &  1.31     & 1.57    \\
Residual   & 1.07 & 1.03  \\
\hline
\end{tabular}
\caption{Signal-to-noise ratios for HLV and HLVIK at design sensitivity, for the total background and the residuals assuming one year of observations.}
\label{SNR_Des}
\end{table}

Adding Kagra and LIGO India to the network, more sources are detected. The reduction of the background is not completely compensated by the improvement of the sensitivity and we observe a small reduction of the SNR: $SNR_{HLV} = 1.07$ and $SNR_{HLVIK} = 1.03$.

Assuming we can confidently claim  a detection with $SNR=3$ (at the level of 3 $\sigma$), we estimate the needed observation time to be 8 (5) years with HLV and its corresponding residual (total) and 8.5 (4) years for HLVIK at design sensitivity. A more conservative $SNR=5$ will require 29(14.5) years with HLV and 23.5(10) years with HLVIK. At that time 3G detectors may be already in operation.

\section{Conclusions}
We have investigated the CBC background derived from a population of isolated binaries 
and from a population of young cluster binaries, including original ($Orig$) and exchanged ($Exch$) binaries. Assuming a fraction of dynamical binaries $f_{Dyn}=0.5$, we find a total background $\Omega_{gw}^{Tot} = 1.2\times{}10^{-9}$, in agreement with previous studies \cite{2008.04890} and with the most recent upper limits derived from GW data \cite{UpperLimits_O3a}. Modelling uncertainties related to 
 star formation and metallicity evolution across cosmic time, we find a possible error of a factor $\sim{} 2$.

The presence of a population of dynamical binaries in young clusters has little effect on the shape of the energy density spectrum in the case of BNSs and BHNSs, but 
it  affects the background from BBHs, adding an extra bump at 100 Hz. This bump, that comes from a specific range of 
redshifted chirp mass (see Fig.\ref{BBHori} and \ref{BBHexc}), is the signature of the population of BBHs from young clusters and depends on the star formation parameters (see Fig.\ref{BBHerr}) and the fraction of dynamical binaries (see Fig.\ref{BBH_prop}). 

The residual background is composed of all the sources that are not detected individually. In the case of 2G detectors, the fraction of detected binaries is small (negligible for BNSs and BHNSs and up to 3\% for BBHs in HLVIK). We have shown that the detected sources, because they are at the lowest redshifts, contribute the most to the total energy density (up to 39\% for BBHs with HLVIK) and subtracting them can  significantly reduce the background amplitude and then affect its detectability. 

We predict that an effective observation time of $\sim{} 8 - 8.5$ years will be necessary to reach a 3~$\sigma$ confidence level with HLV -- HLVIK at design sensitivity.


\paragraph{Discussion}
The CBC background, when detected, can provide valuable information about the star formation history, the metallicity evolution, the mass distribution and the different formation channels. In this study, we explored the impact of adding a population of binaries from young clusters, continuing the work started by  
\cite {2008.04890} for isolated binaries. Other formation channels have been investigated in \cite{Bavera_2021} and confirm that the background shape is deeply affected by the formation/evolution channel of compact binaries.

Our results 
are affected by model uncertainties and parameter choices,  which may have an impact on the population \citep[e.g.,][]{belczynski2022,mapelli2022}. 
For example, we adopted a model in which the natal kicks  are generally low ($\sigma{}=15$ km s$^{-1}$) and depend on the fallback mass \citep{giacobbo2018b}.  Larger kicks ($\sigma{}>100$ km s$^{-1}$) are expected to reduce the number of BNS and BHNS mergers by a factor of ten or more \citep[see][for a discussion]{santoliquido2020b}. Alternatively, models in which the kick does not depend on the fallback lead to a quenching of high-mass BBH mergers.
Our models assume a high accretion efficiency \citep{hurley2002}: during stable mass transfer nearly all the mass lost by the donor is accreted by the companion, unless the latter is a compact object. This assumption is known to have a large impact on the final chirp mass and delay time distribution \citep[e.g.,][]{bouffanais2020}.

We assume a common envelope efficiency $\alpha=5$. This corresponds to an easy ejection of the common envelope with a mild shrinking of the binary semi-major axis. This large value of $\alpha$  is suggested by recent hydrodynamical simulations \citep[e.g.,][]{fragos2019} and by a study of the merger rate \citep[e.g.,][]{santoliquido2020b}, while other works indicate a preference for lower values of $\alpha$ \citep{2021A&A...647A.153B}. Common envelope is certainly one of the main uncertainties in binary evolution models and deserves further consideration.

The evolution of dynamical binaries, especially exchanged systems, is less affected by binary evolution processes (natal kicks, stable mass transfer and common envelope). However, here we consider only one of the possible dynamical formation channels: the evolution of young star clusters. Other families of star clusters are expected to contribute to the overall population of BBHs: globular clusters \citep[e.g.,][]{portegieszwart2000,rodriguez2016,askar2017,rodriguez2018,kremer2020} and nuclear star clusters \citep[e.g.,][]{antonini2016,antonini2019,fragione2020,mapelli2021}. Furthermore, binary compact objects evolving in active galactic nuclei undergo a completely different evolutionary path \citep[e.g.,][]{mckernan2012,mckernan2018,bartos2017,yang2019,tagawa2020,tagawa2021,ishibashi2020}. Also, we neglected the fate of binary compact objects in triple systems, which are characterized by a high-eccentricity sub-population \citep[e.g.,][]{antonini2017,stegmann2022}. We will include the impact of these additional dynamical channels in a follow-up study. 
Nevertheless, the metallicity evolution of the Universe is one of the main uncertainties, because it has a dramatic impact on the population of BBHs (Figure~\ref{BBHerr}).
Hopefully, the increasing number of observations will allow a better understanding and better predictions in the next few years.


\paragraph{Acknowledgments}
We thank the anonymous referee for their useful comments and critical reading of this work. MM, YB, UNDC, NG, CP, SR and FS acknowledge financial support from the European Research Council for the ERC Consolidator grant DEMOBLACK, under contract no. 770017. NG is supported by Leverhulme Trust Grant No. RPG-2019-350 and Royal Society Grant No. RGS-R2-202004.

\bibliography{biblio}
\bibliographystyle{unsrt}
\newpage
\begin{appendix}

\section{Masses distributions histograms}\label{sec:A1}


In this Section, we discuss the origin of the bumps in the BBH background we reported in Section V.a).   
Tables~\ref{tab:orig_stat} and \ref{tab:exch_stat} focus on the main properties of the  sub-populations we show in Figures \ref{BBHori} and \ref{BBHexc} for the $Orig$ and $Exch$ channels, respectively. 
These Tables show that the sub-populations in Figures \ref{BBHori} and \ref{BBHexc} are the result of the interplay between merger redshift and chirp mass. 

\begin{figure}[h!]
\centering
  \includegraphics[width=8cm]{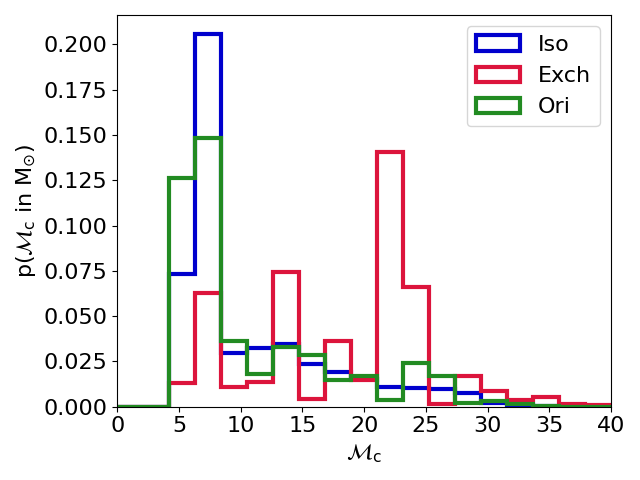} 
\caption{Normalized histogram of the chirp mass for $Iso$ (Blue), $Exch$ (Red) and $Orig$ (Green).}
\label{hist:Mc_z}
\end{figure}

\begin{figure}[h!]
\centering
  \includegraphics[width=8cm]{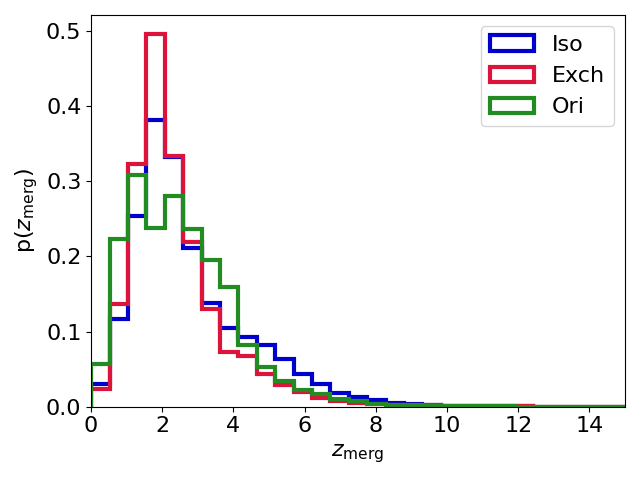}
\caption{Normalized histogram of the redshift for $Iso$ (Blue), $Exch$ (Red) and $Orig$ (Green).}
\label{hist:z}
\end{figure}

Figure\ref{hist:Mc_z} and \ref{hist:z} shows the distributions of chirp mass $\mathcal{M}_{\rm c}$ (left) and merger redshift  $z_{merg}$ (right) for the three different channels studied here ($Iso$, $Orig$ and $Exch$).

\begin{table}[]
\centering
\caption{Quartiles values for the exchanged sub, populations presented in the main text. ({[}0.25,  0.5,  0.75{]}) }
\label{tab:exch_stat}
\begin{tabular}{c|c|c|c|c|c|}
\cline{2-6}
                                                                     & \textbf{All}             & \textbf{{[}$0-65${]}${\rm M}_\odot^{5/3}$}      & \textbf{{[}$65-150${]}${\rm M}_\odot^{5/3}$}    & \textbf{{[}$150-295${]}${\rm M}_\odot^{5/3}$}   & \textbf{{[}$600-1700${]}${\rm M}_\odot^{5/3}$}  \\ 
                                     &             & \textbf{{[}$0-12${]}${\rm M}_\odot$}      & \textbf{{[}$12-20${]}${\rm M}_\odot$}    & \textbf{{[}$20-30${]}${\rm M}_\odot$}   & \textbf{{[}$46-87${]}${\rm M}_\odot$}  \\ \hline
\multicolumn{1}{|c|}{\textbf{$\mathcal{M}_{\rm c} [{\rm M}_\odot]$}} & {[}13.4,  27.7,  21.9{]} & {[}6.0,  6.0,  6.0{]}    & {[}6.6,  6.6,  7.4{]}    & {[}7.1,  7.4,  11.3{]}   & {[}20.9,  21.7,  25.0{]} \\
\multicolumn{1}{|c|}{\textbf{$M\,[{\rm M}_\odot]$}}                  & {[}32.0,  50.0,  51.0{]} & {[}14.1,  14.1,  14.1{]}    & {[}15.5,  15.5,  17.2{]}   & {[}17.0,  17.0,  26.0{]}   & {[}50.0,  50.0,  64.0{]} \\
\multicolumn{1}{|c|}{\textbf{$z_{\rm merg}$}}                        & {[}1.55,  2.05,  5.90{]} & {[}0.60,  0.70,  0.72{]} & {[}1.15,  1.54,  1.76{]} & {[}1.53,  2.54,  2.92{]} & {[}1.60,  1.96,  2.51{]} \\ \hline
\end{tabular}
\end{table}

\begin{table}[]
\centering
\caption{Same as Table~\ref{tab:exch_stat}, but for original binaries.} 
\label{tab:orig_stat}
\begin{tabular}{c|c|c|c|c|c|}
\cline{2-6}
                                                                     & \textbf{All}             & \textbf{{[}$0-65${]}${\rm M}_\odot^{5/3}$}      & \textbf{{[}$65-155${]}${\rm M}_\odot^{5/3}$}    & \textbf{{[}$155-500${]}${\rm M}_\odot^{5/3}$}   & \textbf{{[}$1000-2000${]}${\rm M}_\odot^{5/3}$}  \\ 
&   & \textbf{{[}$0-12${]}${\rm M}_\odot$}      & \textbf{{[}$12-21${]}${\rm M}_\odot$}    & \textbf{{[}$21-42${]}${\rm M}_\odot$}   & \textbf{{[}$63-96${]}${\rm M}_\odot$}  \\ \hline
\multicolumn{1}{|c|}{\textbf{$\mathcal{M}_{\rm c} [{\rm M}_\odot]$}} & {[}6.2,  7.2,  14.4{]}   & {[}5.0,  5.2,  6.6{]}    & {[}5.6,  6.2,  6.7{]}    & {[}6.6,  6.9,  9.0{]}    & {[}14.7,  20.2,  25.2{]} \\
\multicolumn{1}{|c|}{\textbf{$M\,[{\rm M}_\odot]$}}                  & {[}14.3,  16.7,  34.5{]}   & {[}11.4,  11.8,  15.2{]}    & {[}12.9,  14.3,  15.5{]}    & {[}15.2,  16.0,  22.3{]}   & {[}34.4,  48.7,  57.8{]} \\
\multicolumn{1}{|c|}{\textbf{$z_{\rm merg}$}}                        & {[}1.33,  2.32,  3.37{]} & {[}0.58,  0.87,  1.07{]} & {[}1.21,  1.49,  2.01{]} & {[}2.12,  2.97,  3.62{]} & {[}2.23,  2.79,  4.19{]} \\
 \hline
\end{tabular}
\end{table}

By looking at Fig.~\ref{hist:Mc_z},~\ref{hist:z} and at Tables~\ref{tab:orig_stat} and \ref{tab:exch_stat}, we can reconstruct the nature of the sub-populaions. For $Exch$ binaries we extract 4 sub-populations:
\begin{itemize}
    \item \textbf{{[}$0-65${]}${\rm M}_\odot^{5/3}$({[}$0-20${]}${\rm M}_\odot$)}: 
    This sub-population merges at low redshift ($z\sim{0.7}$) from exchanged BBHs with chirp mass $\mathcal{M}_c\sim{6}$~M$_\odot$, which are particularly common in the dynamical simulation \citep[see][]{dicarlo2020b}.
    \item \textbf{{[}$65-150${]}${\rm M}_\odot^{5/3}$({[}$12-20${]}${\rm M}_\odot$)}: This sub-population has the same chirp mass range as the previous one, but merges at higher redshift ($z\sim{1.5}$), close to the peak of cosmic star formation. 
    \item \textbf{{[}$150-295${]}${\rm M}_\odot^{5/3}$({[}$20-30${]}${\rm M}_\odot$)}: This sub-population comes mainly from a population with $\mathcal{M}_{\rm c}\sim {7-12}$M$_\odot$ and $z_{merg}>2$. It is a metal-poor population in the dynamical simulations \citep{dicarlo2020b}. 
    \item \textbf{{[}$600-1700${]}${\rm M}_\odot^{5/3}$({[}$46-87${]}${\rm M}_\odot$)}: This sub-population is associated with the high-mass systems in Fig.~\ref{hist:Mc_z} ($\mathcal{M}_{\rm c}\sim 20-25$M$_\odot$). These systems form from metal-poor binaries ($Z\sim{0.008}$) at redshift $z\sim{2}$ and have a short delay time ($t_{\rm del}<1$ Gyr) in the dynamical simulations.
\end{itemize}
The granularity of the $\mathcal{M}_{\rm c}$ distribution comes from the combination of two factors: the small size of the sample for dynamical binaries (3416 BBH mergers, including exchanged and original binaries) and the fact that \texttt{Cosmo}$\mathcal{R}$\texttt{ate} tends to pick up preferentially the binary systems with a combination of the shortest delay time and the most common metallicity at a given redshift. The combination of these effects results in sharp features in the distribution. The limit to the number of dynamical simulations mainly comes from the computational cost of these simulations: we need $\sim{250}$k GPU hours to obtain a sample of $\sim{200}$ BBH mergers \citep{dicarlo2020b}.

For $Orig$ binaries we extracted 4 sub-populations:
\begin{itemize}
    \item \textbf{{[}$0-65${]}${\rm M}_\odot^{5/3}$({[}$0-12${]}${\rm M}_\odot$)}: This sub-population is composed of low-redshift mergers ($z_{merg}<1.5$) with chirp mass $\mathcal{M}_c\sim{5-7}$~${\rm M}_\odot$ from the peak of the distribution in Fig.~\ref{hist:Mc_z}.
    \item \textbf{{[}$65-155${]}${\rm M}_\odot^{5/3}$({[}$12-21${]}${\rm M}_\odot$)}: This sub-population comes from the first peak in the redshift distribution ($z_{merg}\sim{1.5}$, Fig.~\ref{hist:z}) and chirp mass $\mathcal{M}_c\sim{5-7}$ ${\rm M}_\odot$, Fig.~\ref{hist:Mc_z}. 
    \item \textbf{{[}$155-500${]}${\rm M}_\odot^{5/3}$({[}$21-42${]}${\rm M}_\odot$)}: This sub-population corresponds to the second peak of the redshift distribution ($4>z_{merg}>2$, Fig.~\ref{hist:z}) and chirp mass $\mathcal{M}_{\rm c}\lesssim{10}\,{}{\rm M}_\odot$, Fig.~\ref{hist:Mc_z}.
    \item \textbf{{[}$1000-2000${]}${\rm M}_\odot^{5/3}$({[}$63-96${]}${\rm M}_\odot$)}: This last sub-population comes from high mass ($\mathcal{M}_{\rm c}>10$M$_\odot$) and high redshift mergers ($z_{merg}>2$).
\end{itemize}
The peculiar shape of the redshift distribution is mainly due to the relation between metallicity and redshift. In fact, the $Orig$ sub-population with merger redshift $z_{merg}<2$ mainly originates from stars with metallicity $ Z\sim 0.008$, 
while the one with $z_{merg}>2$ comes from progenitors with lower metallicity ($Z\leq{} 0.002$).

\section{Shared number of sources and energy density proportions in residual background}
\label{App_Prop}
\begin{figure}[h!]
\centering
  \includegraphics[width=17.5cm]{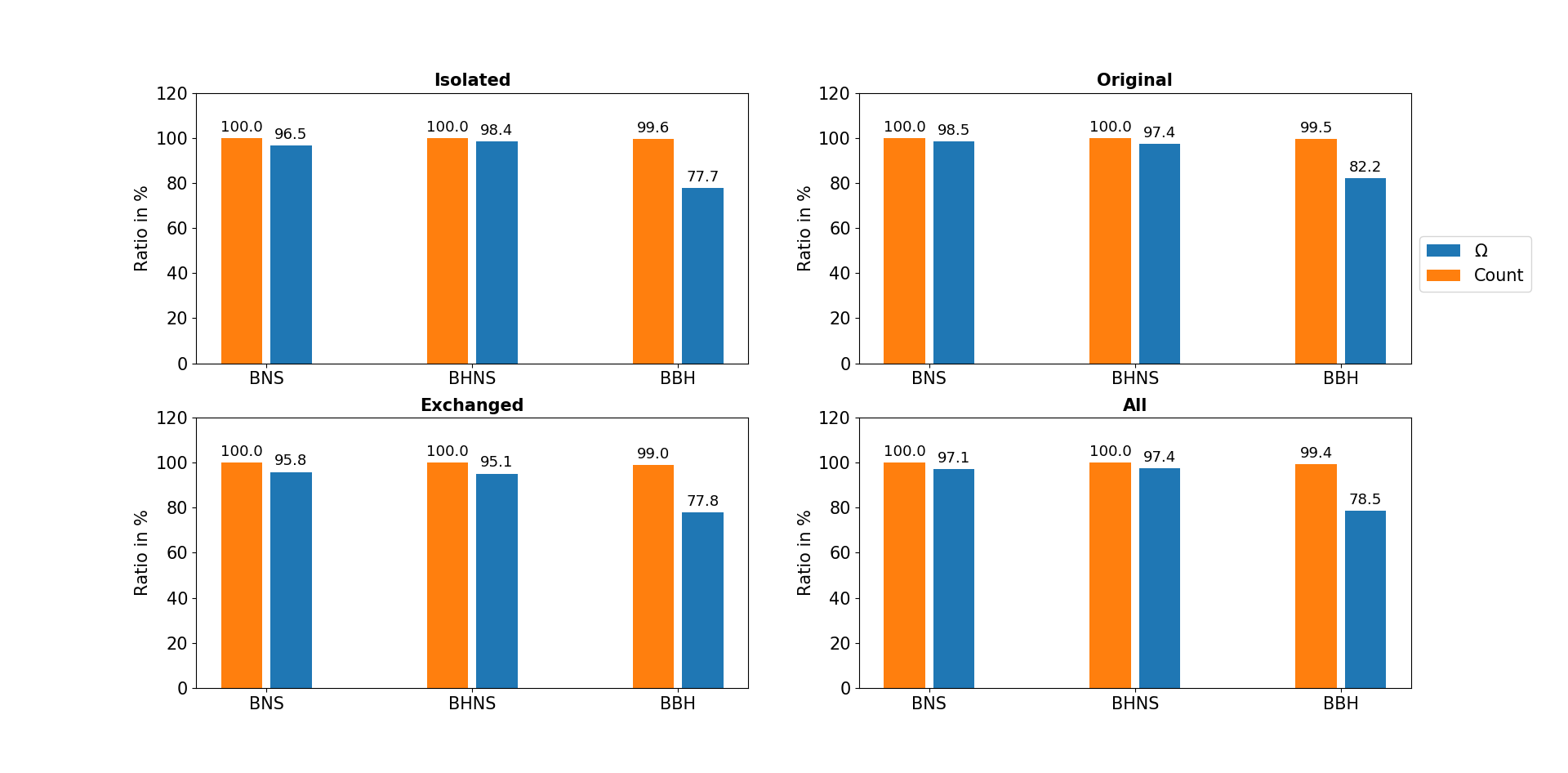} 
\caption{For HLV, ratio of the energy densities of the residual background and the total background (blue), and ratio of the number of sources contributing to the residual background and to the total background (orange), for the different populations (isolated, original, exchanged binaries and the total), from top left to bottom right. }
\label{stat_HLV}
\end{figure}

\begin{figure}[h!]
\centering
  \includegraphics[width=17.5cm]{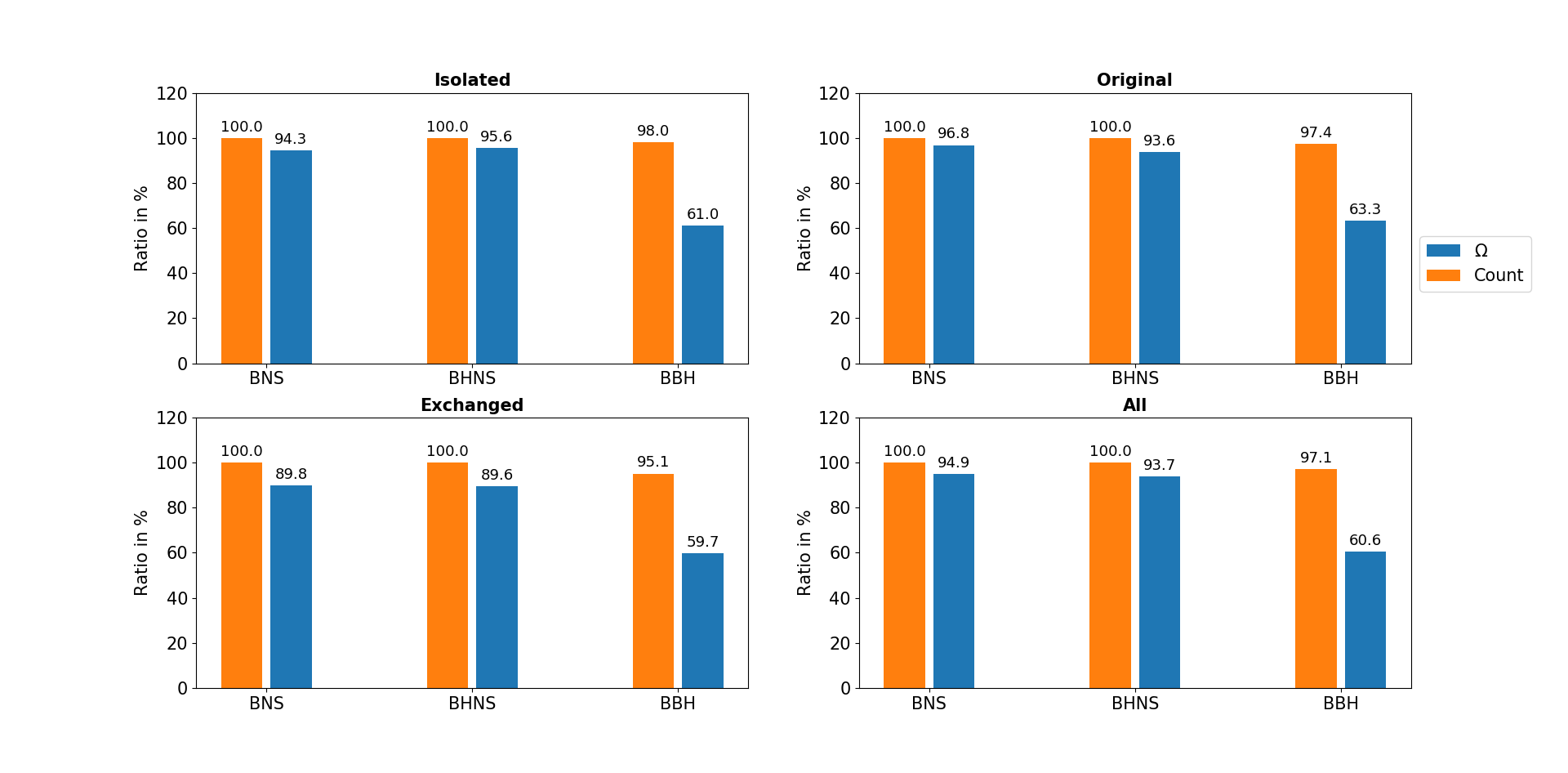} 
\caption{The same as Fig.~\ref{stat_HLV}, but for HLVIK.
}
\label{stat_HLVIK}
\end{figure}

This complementary Section shows the detailed ratios between residual and total backgrounds (HLV, Fig. \ref{stat_HLV}, and HLVIK, Fig.\ref{stat_HLVIK}) for the different formation channels studied here (\textit{Iso, Exch} and \textit{Orig}). 

\end{appendix}
\end{document}